\documentclass{aastex}          %% The manuscript based on AASTeX v5.x
\usepackage{spr-astr-addons}    %% mimicing ASTR journal style
\usepackage{float}

\newcommand{\emaila}{oskari.miettinen@digia.com}

\begin{document}
%% Article title
%
\title{Protostellar classification using supervised machine learning algorithms}

%% Running heads
\shorttitle{Protostellar classification using supervised machine learning}
\shortauthors{Miettinen}

%% Author and Affilations
\author{O.~Miettinen\altaffilmark{1}} 
%\and 
%\author{\altaffilmark{}}
%\affil{Avarea Oy, Rautatiel\"{a}isenkatu 6, FI-00520 Helsinki, Finland}
\email{\emaila} %% non-output

%% Alternate Affilations
\altaffiltext{1}{Digia Plc/Avarea Oy, Rautatiel\"{a}isenkatu 6, FI-00520 Helsinki, Finland}

%\altaffiltext{3}{}

%% Abstract
\begin{abstract}
Classification of young stellar objects\\ (YSOs) into different evolutionary stages helps us to understand the formation 
process of new stars and planetary systems. Such classification has traditionally been based on spectral energy distribution (SED) 
analysis. An alternative approach is provided by supervised machine learning algorithms, which can be trained to classify large 
samples of YSOs much faster than via SED analysis. We attempt to classify a sample of Orion YSOs (the parent sample size 
is 330) into different classes, where each source has already been classified using multiwavelength SED analysis. We used eight 
different learning algorithms to classify the target YSOs, namely a decision tree, random forest, gradient boosting machine (GBM), logistic regression, na\"ive Bayes classifier, $k$-nearest neighbour classifier, support vector machine, and neural network. 
The classifiers were trained and tested by using a 10-fold cross-validation procedure.
As the learning features, we employed ten different continuum flux densities spanning from the near-infrared to submillimetre 
wavebands ($\lambda = 3.6-870$~$\mu$m). With a classification accuracy of 82\% (with respect to the SED-based classes), a GBM algorithm was 
found to exhibit the best performance. The lowest accuracy of 47\% was obtained with a na\"ive Bayes classifier. Our analysis 
suggests that the inclusion of the 3.6~$\mu$m and 24~$\mu$m flux densities is useful to maximise the YSO classification 
accuracy. Although machine learning has the potential to provide a rapid and fairly reliable way to classify YSOs, an SED analysis is still needed to derive the physical properties of the sources (e.g. dust temperature and mass), and to create the labelled training data. The machine learning classification 
accuracies can be improved with respect to the present results by using larger data sets, more detailed missing value imputation, and advanced ensemble methods (e.g. extreme gradient boosting). Overall, the application of machine learning 
is expected to be very useful in the era of big astronomical data, for example to quickly assemble interesting 
target source samples for follow-up studies.  
\end{abstract}

%% Keywords

%%  Please use labels (\label, \ref) for section, figure, table, 
%%  equation  reference. Use \cite for bibliography references.
%
%\section{}%\label{s:?}
%\subsection{}%\label{ss:?}
%\subsubsection{}%\label{sss:?}

%% Math 
%
%\begin{eqnarray}%\label{eqn:?}
%\\ \nonumber
%\end{eqnarray}
%
%\begin{equation}%\label{eqn:?}
%\end{equation}

\section{Introduction}%\label{s:?}

An essential part of the star formation studies is to try to classify
the young stellar objects (YSOs) into different evolutionary stages, and construct
a coherent YSO evolutionary sequence. Also, by determining the relative percentages of YSOs in different stages, the statistical time spent in each stage can be constrained, which in turn helps to quantify the overall timescale of the stellar birth process in different molecular cloud environments
(e.g. \citealp{evans2009}; \citealp{dunham2015}). 

Considering the formation of low-mass, solar-type stars, the YSOs have traditionally been classified 
into distinct stages on the basis of their infrared (IR) spectral slopes (e.g. \citealp{lada1984}) or 
bolometric temperatures (\citealp{myers1993}). In particular, the spectral energy distribution (SED) of a YSO, 
which is characterised by the bolometric temperature and luminosity, is commonly used to determine the evolutionary 
stage of the source, that is whether it is a so-called Class~0 or I protostar, or Class~II or III 
pre-main sequence (PMS) star (e.g. \citealp{lada1987}; \citealp{adams1987}; \citealp{andre1993}; see also \citealp{andre2000} for a review). 
Indeed, an SED analysis is very useful, not just for the purpose of source classification, 
but to derive some of the key physical properties of the source, such as the dust temperature and dust mass. 
However, modelling the source SEDs can be fairly time consuming, and hence, 
to quickly determine the evolutionary classes for a large sample of YSOs, an automated procedure 
that employs the observed source properties (i.e. the flux densities) would be very useful. 
In this regard, machine learning has the potential to yield a fast way to classify sources  
(as compared to an SED analysis) as long as the algorithm(s) in question can be trained with data sets composed 
of relevant flux densities and corresponding evolutionary classes of the target YSOs. 

So far, machine learning based classification of astrophysical objects has mostly been applied in extragalactic research 
(e.g. \citealp{krakowski2016}; Aniyan \& Thorat 2017; \citealp{sreejith2018}; \citealp{beck2018}; Pashchenko et al. 2018; 
\citealp{hui2018}; Lukic et al. 2018; \citealp{an2018}; see also \citealp{lochner2016}), 
while Galactic machine learning studies have been relatively few in number (e.g. \citealp{marton2016}; \citealp{yan2018}). 
Hence, pilot studies about using machine learning in YSO classification, which the present work represents, are warranted.

In this paper, we report the results of our protostellar classification test using 
several different supervised machine learning algorithms. The data set used in
this study is described in Sect.~2, while the data analysis is presented in Sect.~3. The results are presented 
and discussed in Sect.~4, and in Sect.~5 we summarise the key results and conclusions of this work.

\section{Data}

The data analysed in this paper were taken from Furlan et al. (2016, hereafter FFA16). As part of 
the \textit{Herschel}\footnote{\textit{Herschel} is an ESA space observatory with science instruments provided 
by European-led Principal Investigator consortia and with important participation from NASA.} Orion Protostar Survey 
(HOPS; e.g. \citealp{stutz2013}), FFA16 studied and modelled the SEDs of a large, homogeneous sample of 
330 YSOs in the Orion molecular cloud complex (the authors assumed a uniform distance of 420~pc to the cloud complex). 
At the time of writing, this is the largest available YSO sample 
investigated in a single star-forming cloud complex. The photometric data employed by the authors included the $J$, $H$, 
and $K_{\rm S}$ near-IR data from the Two Micron All Sky Survey (2MASS; \citealp{skrutskie2006}), \textit{Spitzer} 
IR data obtained with the Infrared Array Camera (IRAC; 3.6--8.0~$\mu$m; \citealp{fazio2004}),  
the Multiband Imaging Photometer for \textit{Spitzer} (MIPS; 24~$\mu$m; \citealp{rieke2004}), and 
the Infrared Spectrograph (IRS; 5.4--35~$\mu$m; \citealp{houck2004}). The \textit{Herschel} satellite (\citealp{pilbratt2010}) 
was used to observe the 70, 100, and 160~$\mu$m far-IR bands with the Photodetector Array Camera and Spectrometer 
(PACS; \citealp{poglitsch2010}), while the 350 and 870~$\mu$m submillimetre data were obtained with the   
Atacama Pathfinder EXperiment (APEX; \citealp{gusten2006}) using its Submillimetre APEX
BOlometer CAmera (SABOCA; Siringo et al. 2010) and Large APEX BOlometer CAmera (LABOCA; \citealp{siringo2009}). 
The aforementioned bands cover the typical protostellar SED peak emission at $\sim100$~$\mu$m and its surrounding wavelengths, which 
is essential to determine reliable physical properties and evolutionary stage of the source (see below).

On the basis of their panchromatic SED analysis, FFA16 classified their YSO sample into
92 Class~0 protostars, 125 Class~I protostars, 102 flat-spectrum sources (expected to be objects in transition between 
the Class~I and II phases), and 11 Class~II PMS stars. The corresponding relative percentages are 
$27.9\% \pm 2.9\%$, $37.9\% \pm 3.4\%$, $30.9\% \pm 3.1\%$, and 
$3.3\% \pm 1.0\%$, respectively, where the quoted uncertainties represent the Poisson counting errors. 
We note that one of the FFA16 sources, namely the Class~0 source HOPS~400, was first uncovered by Miettinen et al. 
(2009; their source SMM~3) in their LABOCA imaging of the Orion~B9 star-forming region (see also \citealp{miettinen2016}, 
and references therein).  

In brief, the physical explanation of why the SED analysis presented by FFA16 can be used to classify YSOs into 
different evolutionary stages is as follows (see e.g. \citealp{white2007}; \citealp{dunham2014} for reviews). The youngest protostellar objects, or Class~0 objects, are characterised 
by a central protostar deeply embedded in its cold, dusty envelope (\citealp{andre1993}; \citealp{andre1994}). Hence, the source 
is extremely faint in the optical ($\lambda \sim0.4-0.7$~$\mu$m) and near-IR ($\lambda \gtrsim0.7-5$~$\mu$m; traced by 
2MASS and \textit{Spitzer}/IRAC observations), but bright in the far-IR ($\lambda \sim25-350$~$\mu$m; traced by \textit{Spitzer}/MIPS 
and \textit{Herschel} observations) and (sub-)millimetre ($\lambda \gtrsim350-1\,000$~$\mu$m; traced by APEX bolometer 
observations) dust emission. The central protostar 
increases its mass by accreting gas from the surrounding envelope via a circumstellar disk. When the envelope mass has dropped 
to that of the growing central protostar, the system is believed to transition from the Class~0 to the Class~I stage. Class~I protostars 
are still surrounded by an accretion disk and a circumstellar envelope of gas and dust, and hence their SEDs peak in the far-IR. 
However, another prominent SED bump can be seen in the mid-IR ($\lambda \sim5-25$~$\mu$m; traced by 
\textit{Spitzer}/IRAC and MIPS observations), which is an indication of hotter dust than in the previous Class~0 stage. 
If the central protostar can be seen along the long axis of the protostellar outflow, the Class~I object can be optically visible. 
The intermediate stage between the Class~I and Class~II stages is characterised by the disappearing dust excess emission in the mid-IR, 
and hence the sources in this transition stage are known as flat-spectrum (SED) sources (\citealp{greene1994}).
In the Class~II stage, the envelope has dissipated, and an optically visible PMS star is surrounded by a tenuous disk. The SEDs of
Class~II objects peak at visible or near-IR wavelengths, and the disk adds an IR excess to the SED.

\section{Data analysis}

\subsection{Source selection}

One of the four YSO classes from FFA16, namely the Class~II phase, 
is highly unbalanced with respect to the other three classes (Class~0, Class~I, and flat sources). 
Indeed, only $3.3\%$ of the FFA16 sources were classified as Class~II sources, 
which is about ten times less than the occurrence of other types of sources. 
Although techniques to deal with imbalanced data sets exist (e.g. \citealp{kotsiantis2006a} for 
a review; \citealp{he2013}), we do not consider the FFA16 Class~II sources in the subsequent analysis because their 
relative rarity can lead to problems in the training of the supervised classification algorithms, 
and in the evaluation of the classifiers' performance.

After discarding the Class~II sources, we are left with 319 sources, out of which $28.8\% \pm 3.0\%$ are 
Class~0 sources, $39.1\% \pm 3.5\%$ are Class~I sources, and $32.0\% \pm 3.3\%$ are flat-spectrum sources. 
Hence, all these three classes are in good relative number balance with respect to each other.

\subsection{Missing value treatment}

When developing machine learning models, it is important to handle the missing values 
in the analysed data set (e.g. \citealp{witten2005}; Saar-Tsechansky \& Provost 2007). 
A common approach is to replace the missing values by the mean of 
the non-missing values for the variable or feature in question (e.g. \citealp{alpaydin2010}; \citealp{lantz2015}). 
This imputation method does not change the sample mean of the variable.

We found that $69.3\%$, $54.9\%$, and $40.1\%$ of the selected FFA16 sources missed 
the 2MASS $J$, $H$, and $K_{\rm S}$ near-IR data. Owing to these large percentages, we discarded the 
2MASS data in our subsequent analysis. Also, $31\%$ of the selected FFA16 YSOs lacked the SABOCA 350~$\mu$m data, 
but we included this waveband in the analysis because it provides a useful data point in 
the Rayleigh-Jeans part of the source SED, and the band has also been covered by observations with other instruments, 
such as by \textit{Herschel} submm surveys (e.g. \citealp{andre2010}). All the remaining bands except 
the \textit{Herschel} 70~$\mu$m band were also found to contain missing values, but in those cases only 0.3\% to 
16.6\% (11\% on average) of the sources had missing data. 

To impute the missing values, we used the {\tt R} package {\tt MICE} (Multivariate Imputation via Chained Equations; 
\citealp{vanbuuren2011}). The usage of {\tt MICE} is based on the assumption that the missing data are Missing at Random (MAR), 
and it imputes data on a variable by variable basis (i.e. flux density by flux density basis in our case) by specifying 
an imputation model per variable. To calculate the imputations, we used the predictive mean matching (PMM) method 
(\citealp{little1988}). If $S_{\rm miss}$ is the variable that contains missing data, and $S_{\rm i}$ are 
the variables that do not suffer from missing data, the PMM algorithm works as follows: 
i) for cases with no missing data, a linear regression model of $S_{\rm miss}$ is estimated on $S_{\rm i}$, which yields a set of coefficients 
$b$; ii) a random draw is taken from the distribution of $b$ values, which yields a new sample of coefficients $b^*$; this step 
is needed to generate some random variability in the imputed values; iii) the $b$ values are used to calculate the predicted values 
for the observed $S_{\rm miss}$ values, while the $b^*$ values are used to calculate the predicted values 
for the missing $S_{\rm miss}$ values; iv) for each case with 
a missing $S_{\rm miss}$ value, a set of cases with $S_{\rm miss}$ present is identified where the predicted values are closest to 
the predicted value for the case with a missing $S_{\rm miss}$ value; v) from the latter cases, a random value is chosen, and it is then
used to impute the missing value. The PMM method is expected to lead to more reasonable estimates of the missing flux densities 
than simply using the sample averages, while still being a very fast imputation method.

Regarding the traditonal mean or median imputations, one might think that replacing the missing flux density values by the 
mean or median values for each protostellar class separately would be a better approach than using the full sample mean or median. 
Indeed, this would be a more physical approach than using the full sample mean or median imputation because 
the flux density at a given wavelength can evolve as the source evolves (e.g. the amount of dust in the protostellar envelope 
decreases as the source matures, which affects the far-IR and submm emission considered in the present study). However, to do this, 
one would have to use information from the test set's labels or classes, 
and one should not leak such information (which is correlated with the labels) to the training procedure of a classifier. 
More importantly, when analysing new, previously unseen data, the protostellar or YSO classes are not even known, 
but they are what one wants to determine or predict.

On the other hand, it is known that for example 
the far-IR and submm flux densities considered in the present work 
depend on each other via the frequency dependent dust opacity, $\kappa_{\nu} \propto \nu^{\beta}$, where 
$\beta$ is the dust emissivity index (e.g. \citealp{shetty2009}). Hence, the missing submm flux densities could 
also be estimated in a source-by-source fashion 
from the existing ones by assuming a value for $\beta$. However, estimating the missing flux density values this way 
requires more feature value engineering, and hence is not as fast as for example the PMM. Indeed, obtaining 
fast classifications (as compared to an SED analysis) is the key of applying machine learning in the first place. 

\subsection{Training and test data sets}

A supervised machine learning algorithm requires a so-called training set through which the algorithm tries to 
learn how the input values, or features (flux densities in our case), map to the response values, 
or labels (protostellar classes in the present work). Another data set, the so-called test set, is then used 
to test the performance of the trained model by showing it only the input values, and see which corresponding 
classes the model predicts for the data it did not see during training.

A traditional way to create the training and test data sets is to fragment the original data into two parts. 
For example, one common way is to use 80\% of the data for training the algorithm, and the remaining 20\% for testing 
it (the so-called 80/20 rule or the Pareto principle; e.g. \citealp{box1986}). However, the present data set is fairly small 
for a machine learning experiment (319 sources in total), and hence the aforementioned data splitting would not yield good training 
and test capabilities. Instead, we used the technique of $k$-fold cross-validation (CV), where the data set is randomly divided 
into $k$ roughly equal sized subsamples, or folds (\citealp{mosteller1968}; see e.g. \citealp{james2017}, Sect.~5.1.3 therein). 
The algorithm is trained using $k-1$ of the folds, and the resulting model is tested on the remaining part of the data. 
This procedure is then repeated $k$ times. We did the sampling with replacement, which means that the same source could be 
sampled more than once. The final prediction performance was taken as the mean of the $k$ results. 
A value of $k=10$ was used in the present work. 

Another caveat of dealing with small data sets in machine learning is the problem of overfitting, which 
means that an algorithm starts to learn the details of the training data set, including the random noise features, such 
as outliers instead of (or besides) the underlying general rules, patterns, or relationships (e.g. \citealp{hawkins2004}). 
This weakens the algorithm's ability to generalise well to previously unseen cases, because the aforementioned noise features are 
unlikely to occur in the test and new data sets. However, the usage of $k$-fold CV can reduce the degree of overfitting (but not fully prevent it) 
owing to the split of the data into multiple training and test sets (e.g. \citealp{cawley2010}).

\subsection{Principal component analysis}

Another data preprocessing step we did was principal component analysis 
(PCA; \citealp{pearson1901}; Hotelling 1933; \citealp{jolliffe2002}; Abdi \& Williams 2010), 
which is a common technique in machine learning for extracting the most important variables among a large number of 
variables in a data set, and hence to overcome feature redundancy and reduce the dimensionality of the problem. 
We note that if the analysed data set is split into separate
labelled training set and test set, then the PCs need to be calculated on the training data set. However, 
because in the present study we employed the method of $k$-fold CV for training and testing the classifiers, 
we carried out the PCA using the full data set (i.e. 319 sources). Because in the PCA the original data are 
projected onto directions that maximise the variance, the variables were scaled to have a variance equal to 
$\sigma_{\rm SD}^2=1$, where $\sigma_{\rm SD}$ is the standard deviation, before the PCs were calculated.

\begin{figure}[H]
\centering
\resizebox{\hsize}{!}{\includegraphics{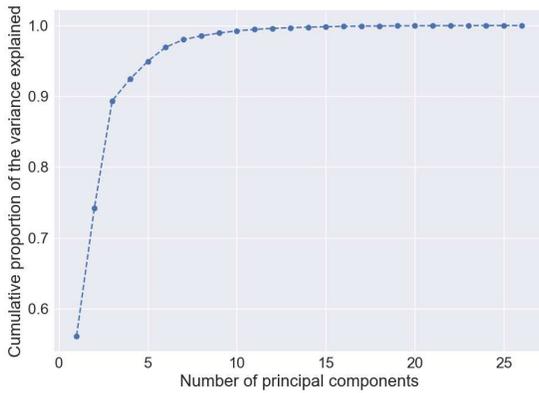}}
\caption{Cumulative proportion of the variance in the data explained by each PC. The first ten PCs explain about 99.2\% of 
the variance.}
\label{figure:pca}
\end{figure}

\begin{figure*}%[H]
\begin{center}
\includegraphics[width=0.97\textwidth]{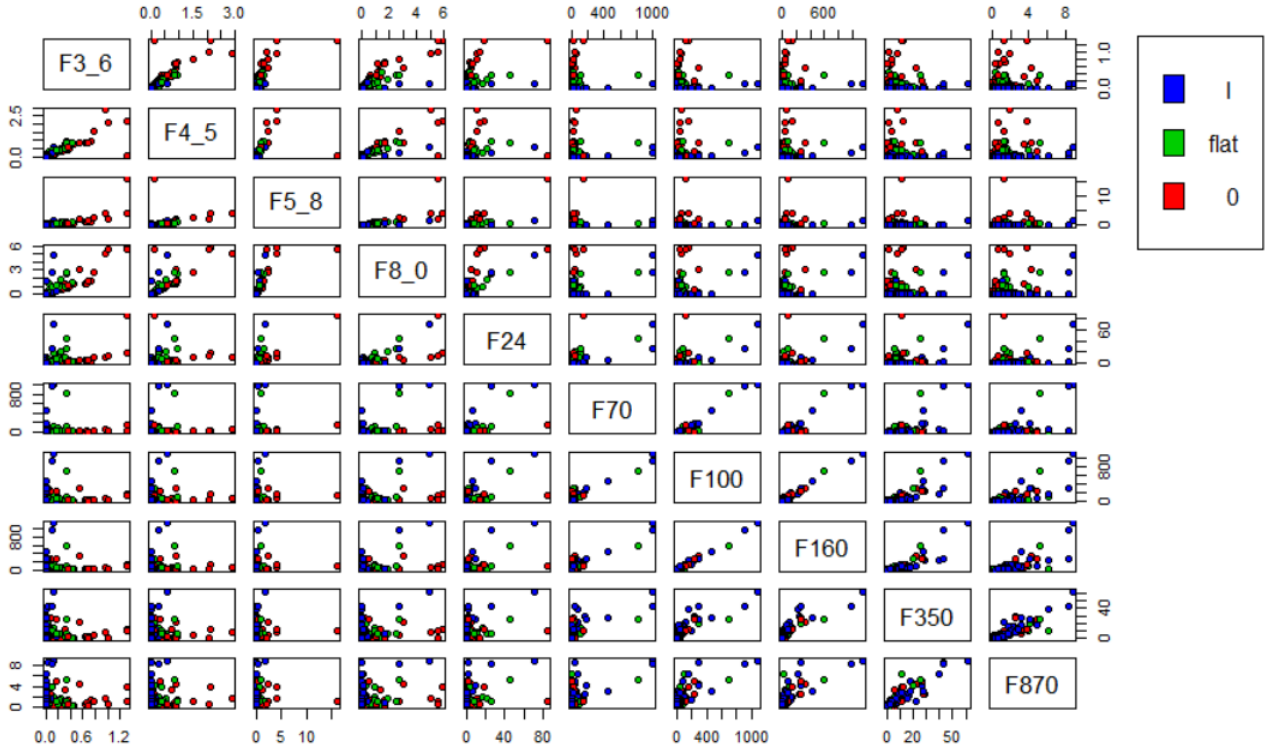}
\caption{Draftsman's plot showing the values of each of the considered continuum flux densities (in Jy) against each other.}
\label{figure:scatter}
\end{center}
\end{figure*}

In Fig.~\ref{figure:pca}, we plot the cumulative proportion of the variance in the data set explained by each PC. 
In this analysis, all the other wavebands except those of 2MASS were taken into account (i.e. four \textit{Spitzer}/IRAC 
bands, one \textit{Spitzer}/MIPS band, three \textit{Herschel}/PACS bands, two APEX bands, and 16 \textit{Spitzer}/IRS bands).
We found that the first ten PCs explain about 99.2\% of the variance in the data, and hence we used only the 
first ten features in the subsequent analysis. These ten features correspond to the continuum flux densities 
in the FFA16 photometric data, that is from the \textit{Spitzer}/IRAC 3.6~$\mu$m to LABOCA 870~$\mu$m data, 
while the \textit{Spitzer}/IRS data were discarded. 
The distributions of the considered flux densities are presented in Fig.~\ref{figure:scatter}, which shows a draftsman's plot, 
or a scatter matrix plot (see e.g. NIST/SEMATECH e-Handbook of Statistical 
Methods\footnote{{\tt https://www.itl.nist.gov/div898/handbook/eda/section3/\\scatterb.htm}}), 
and which enables to see the interrelations between variables in multivariate data (ten 
dimensional in our case). The plot consists of an array of two-variable scatter diagrams. 
For example, the plot demonstrates the strong correlation between the \textit{Herschel} 100~$\mu$m and 160~$\mu$m 
flux densities.

Besides reducing the dimensionality of the problem, the feature selection enabled by PCA also helps to relieve 
the problem of overfitting discussed in Sect.~3.3. The reason for this is that the trained model will be less complex 
the smaller the number of features is relative to the number of data cases (or rows). Hence, although the present data set is 
fairly small and hence subject to overfitting, both the $k$-fold CV and PCA employed in the present study can alleviate 
the influence of overfitting.

\subsection{Machine learning classification}

After data preparation, the following eight supervised classifiers were tested on the FFA16 data: a decision tree (e.g. 
\citealp{quinlan1986}; \citealp{murthy1998}), random forest (\citealp{ho1995}; \citealp{breinman2001}), gradient boosting machine 
(GBM; \citealp{breinman1997}; \citealp{friedman2001}), logistic regression (\citealp{cox1958}), na\"ive Bayes classifier 
(e.g. \citealp{domingos1996}; \citealp{mccallum1998}; \citealp{zhang2004}), $k$-nearest neighbours ($k$-NN; e.g. 
Cover \& Hart 1967; \citealp{altman1992}), support vector machine (SVM; e.g. \citealp{vapnik1963}; Cortes \& Vapnik 1995; 
Burges 1998; \citealp{christianini2000}; Scholkopf \& Smola 2001), and artificial neural network (e.g. McCulloch \& Pitss 1943; 
\citealp{rosenblatt1958}; Jeffrey \& Rosner 1986; \citealp{zhang2000}). We refer to Kotsiantis et al. (2006b), Kotsiantis (2007), and Ball \& Brunner (2010) for reviews of the aforementioned algorithms. As mentioned in Sect.~3.3, the classifiers were trained 
and their performance was tested using the technique of 10-fold CV. In what follows, is a brief description of 
each of the tested classifier.

\subsubsection{Decision tree}

A decision tree classifier attempts to learn simple decision rules that can predict the label for an instance 
on the basis of its feature values. The data are split according to the feature values in the so-called decision nodes, 
and the final leaf nodes contain the outputs (i.e. the protostellar classes in our case). One caveat of decision trees is that they are subject 
to overfitting if too complex trees are being built, and hence they might not generalise well.

To buid a decision tree classifier, we used the {\tt R} package {\tt rpart} 
(Recursive Partitioning and Regression Trees), which uses the Classification and Regression Trees (CART; Breinman et al. 1984) 
algorithm. The CART algorithm employs binary splits on the input variables to grow the tree, and the splits were evaluted 
on the basis of the Gini index (a Gini score of zero means a perfect separation). 

\subsubsection{Random forest}

Contrary to a single tree CART model, random forest 
takes random subsamples of both the observations and features from the training data (bagging), 
and trains decision trees on those cases. A whole army of such decision trees are grown, and the most 
common outcome for each observation is used as the final result. Such approach enable random forests to limit 
overfitting, which makes them very powerful classifiers. 

For the purpose of random forest classification, we employed 
the {\tt R} package {\tt randomForest}, which implements Breinman's random forest algorithm. As the number of trees, we used the 
{\tt randomForest}'s default value of $n_{\rm tree}=500$, and the subsamples were chosen randomly with replacement. 
The number of variables that were randomly sampled was set at $\sqrt{p}$, where $p$ is the total number of variables in the original 
data set ($p=10$ in our case). The winning class was defined as the one with the highest ratio of proportion of votes to the cutoff 
parameter, where the latter was set at $1/k$, where $k$ is the number of classes ($k=3$ in our case). The minimum terminal 
node size, which controls the depth of the tree (the larger the parameter is, the smaller the tree will be), was set at unity, 
while in terms of the number of terminal nodes, the trees were let to grow to the maximum possible size.

\subsubsection{Gradient boosting}

While random forest is a bagging method with trees being run in parallel 
and without interaction, gradient boosting is an ensemble method where decision trees are added 
to learn the misclassification errors in existing models, and this sequentially boosts the training procedure. Because 
gradient boosting is a greedy algorithm (i.e. it makes the optimal choice at each step (locally optimal choice) as it 
tries to reach the overall optimal way to solve the classification problem), it can quickly overfit the training data set.

The boosting was performed using the {\tt R}'s {\tt gbm} algorithm, whose implementation follows the 
Friedman's GBM (\citealp{friedman2001}). Because our classification problem 
is composed of more than two classes, the analysis was carried out as a multinomial version. As the metric of 
the information retrieval measure, we used the normalised discounted cumulative gain ({\tt metric = ndcg}).
The total number of trees to fit, which corresponds to the number of iterations, was set at $n_{\rm tree}=500$ 
as in the case of our random forest classification. The value of $n_{\rm tree}$ also corresponds to the 
number of basis functions that are being iteratively added in the boosting process (each additional basis function further 
reduces the loss function). The interaction depth, which represents the maximum depth of variable interactions, was fixed at $k=3$. 
Hence, the number of terminal nodes or leaves, which is given by $J=k+1$, was set at $J=4$ (for comparison, $J=2$, or a decision stump, 
means that no interactions between variables is allowed). 
The shrinkage parameter ($0<\nu\leq1$), which represent the learning rate, 
was set at $\nu=0.005$. High learning rates of $\nu\simeq1$ ($\nu=1$ means no shrinking) are expected to result in overfit models, 
while small shrinkage parameter 
values ($\nu\leq0.1$) slow down the learning process, but are expected to lead to much lower generalisation error (\citealp{friedman1999}). 
Finally, the minimum number of observations in the trees' terminal nodes was set to unity. Such small value was chosen because our training samples were so small.

\subsubsection{Logistic regression}

Despite its name, logistic regression is a classification algorithm rather than a regression technique.
Logistic regression uses a logistic function and the predictor feature values to model the probabilities for 
an instance to belong to different classes.

To estimate a multinomial logistic regression model, we used the {\tt R} algorithm called {\tt multinom}. The algorithm predicted the probabilities for each source to belong to the three different classes (Class~0, Class~I, and flat-spectrum sources), 
and the final assignment was done according to the highest predicted probability.

\subsubsection{Na\"ive Bayes}

Similarly to logistic regression, na\"ive Bayes classifiers belong to a family of probabilistic classifiers. 
Here, the classification relies on the Bayes' Theorem under the na\"ive
assumption that the features are independent of each other.

To calculate the conditional posterior probabilities for 
our categorical protostellar class variable given the flux densities as predictor variables, we used the {\tt naiveBayes} 
algorithm of {\tt R}. No Laplace smoothing was applied, which would prevent the frequency-based probability estimate to be equal 
to zero as a result of a given class and feature value never occuring together in the training data. The latter is not 
an issue for the present data set, because the missing feature values were imputed.

\subsubsection{$k$-nearest neighbours}

The $k$-NN algorithm is a member of the so-called instance-based, lazy-learning algorithms (Mitchell 1997). 
Here, a new, unseen instance is classified by comparing it with  
those $k$ training cases  that are closest in feature space (i.e. have similar properties with the new case), 
and the new case's class is then determined by a majority vote of its neighbours' classes.
To find these $k$-nearest neighbours, the value of $k$ needs to be specified, and a distance metric is required. 

\begin{figure}[!htb]
\centering
\resizebox{\hsize}{!}{\includegraphics{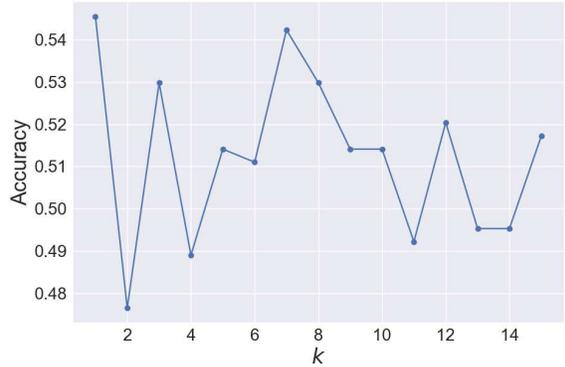}}
\caption{Accuracy of the $k$-NN classifier as a function of the number of nearest neighbours. The best peformance was reached when $k=1$, but we selected the next best performace reached with $k=7$ because the 1-NN classifier is subject to 
overfitting (see text for details). }
\label{figure:acc_knn}
\end{figure}

To carry out a $k$-NN classification, we used the {\tt R}'s {\tt knn3} algorithm. 
We run our $k$-NN classification by experimenting with different values of $k$, ranging from $k=1$ to $k=15$, 
and by adopting the Euclidean distance metric. As shown in Fig.~\ref{figure:acc_knn}, 
the best classification performance was reached when $k=1$ (considering only the closest neighbour).  
The $y$-axis of Fig.~\ref{figure:acc_knn} shows the overall accuracy of the classification, which is defined as 
the ratio of cases that are correctly classified to the total number of cases. However, a 1-NN classifier 
can lead to overfitting and does not generalise well enough to other YSO samples (a small $k$
means that noise has a higher effect on the classification). Hence, we consider 
the next highest accuracy that was reached when $k=7$ in our subsequent comparison of different classifiers.

\subsubsection{Support vector machine}

The basic SVMs are binary classifiers, where the 
idea is to find the dividing hyperplane that both separates the two classes in the 
training data set and maximises the margin between the boundary members, that is between the SVs.
A new instance is classified by examining on which side of the hyperplane it falls. 
In the case of non-linear classification, the goal 
is to find a hyperplane that is a non-linear function of the input variables. 
This is done by the so-called kernel trick, where the input features are mapped into 
a higher dimensional feature space. Besides binary classification, SVMs can also perform multiclass 
classification, which is the case in the present study ($k=3$ classes). There are various options to do that, 
and we used a balanced one-against-one classification strategy, where three binary classifiers 
were trained (the number of classifiers is given by $k(k-1)/2$), and a simple voting strategy among them was applied 
to classify a new instance. 

Our SVM classification was done using the {\tt ksvm} algorithm of {\tt R}, 
and to create a non-linear classifier we used a Gaussian radial basis function (RBF) kernel. The algorithm was 
set to calculate the inverse kernel width for the RBF directly from the data (rather than specifying its value). 
The value of the regularisation constant $C$ was set equal to unity, where the effect of $C$ is such that 
the larger it is, the narrower the margin between the SVs is, and hence the classifier is more prone 
to overfitting. A smaller $C$ means a wider margin, and hence more misclassifications in the training set, which 
in turn can lead to underfitting issues.

\subsubsection{Neural network}

A neural network classifier reads in the input features in the so-called input layer,
and, in the case of a multi-layer perceptron, attempts to learn a non-linear function approximator 
to correctly classify the training cases' target variables appearing in the so-called output layer.

To build a neural network classifier, we used the {\tt R} package {\tt nnet}, 
which fits a single-hidden-layer neural network, that is there is only one hidden layer between the aforementioned input 
and output layers. The feature data are being weighted, and transfered to the 
nodes or neurons in the hidden layer. The hidden layer neurons 
process the sum of the weighted inputs by applying a so-called transfer function, 
and pass the results forward. In the present work, we adopted a sigmoid shaped logistic transfer function, which is appropriate 
for discrete outputs such as protostellar classes.
The maximum number of iterations used was set to 100, and the weight decay in the weight update 
rule was set equal to zero, which means that the weights did not exponentially decay to zero in case of no other updates were 
being scheduled (during the training phase, the update steps modify the weights applied on the input features).

\begin{figure*}%[H]
\begin{center}
\includegraphics[width=0.47\textwidth]{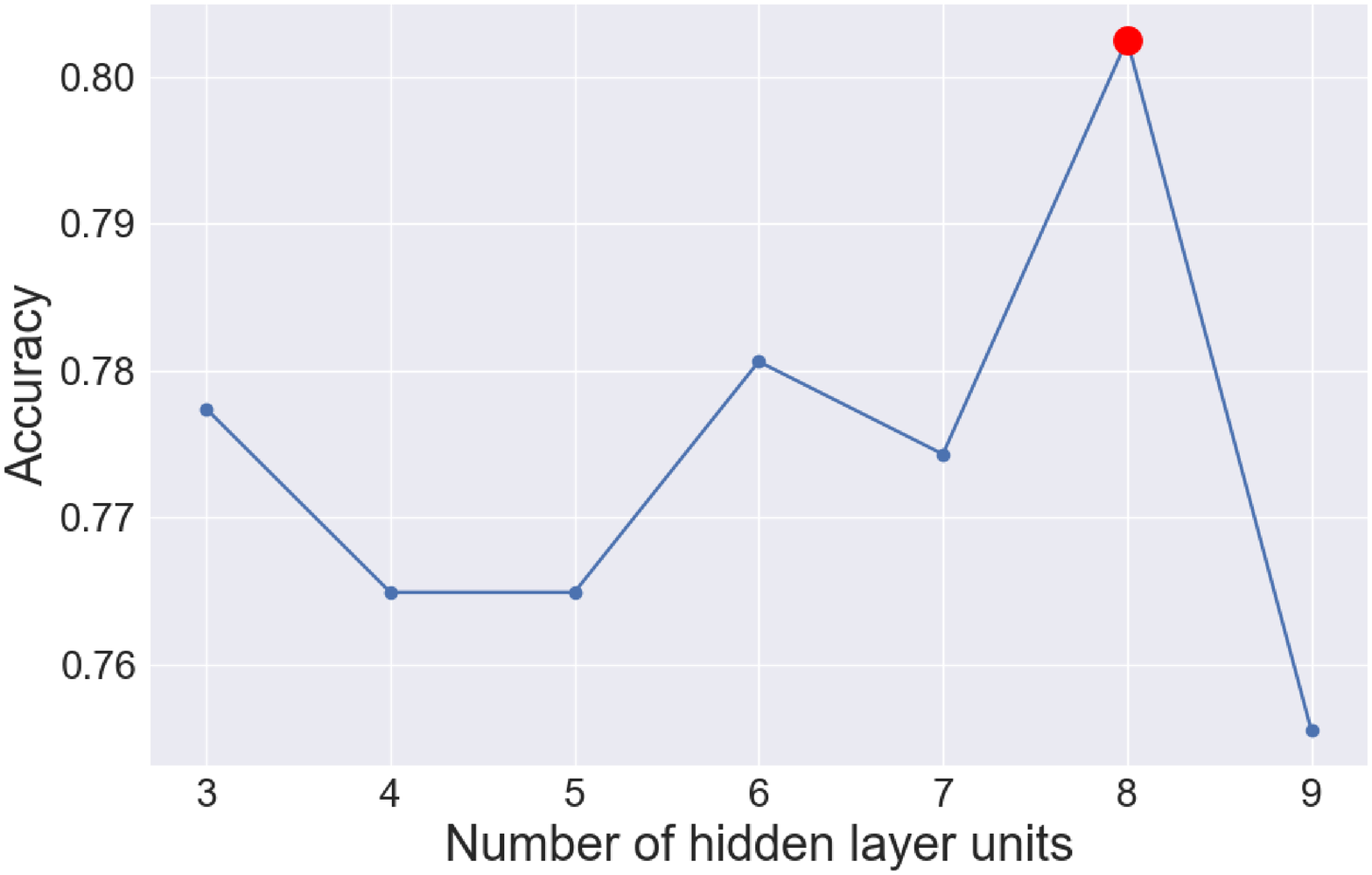}
\includegraphics[width=0.47\textwidth]{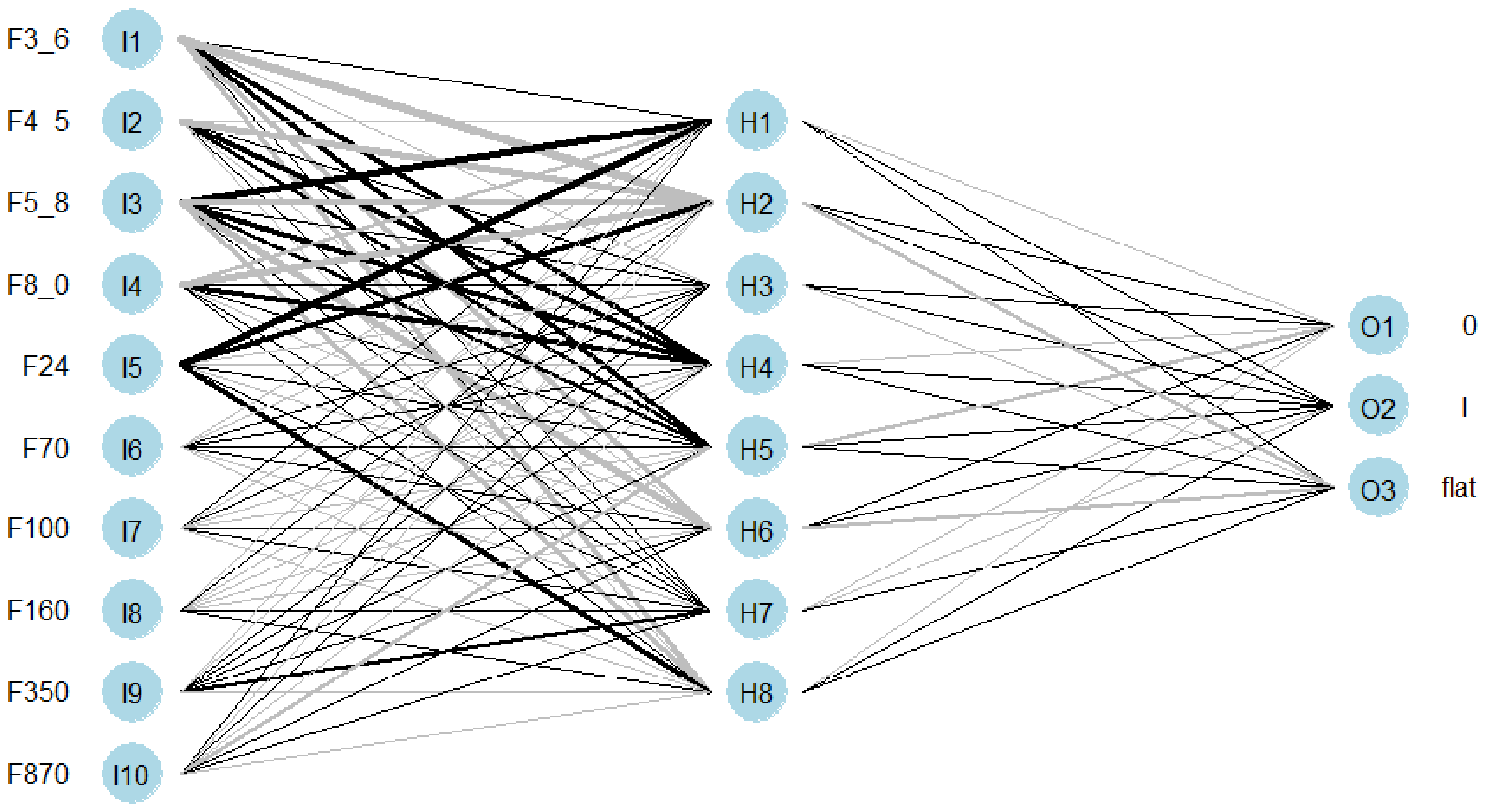}
\caption{\textbf{Left:} Accuracy of the single-layer neural network classifier as a function of the number of 
neurons or nodes in the hidden layer. The red, filled circles indicates the highest accuracy, which was obtained with eight nodes. 
\textbf{Right:} Artifical neural net comprised of eight hidden layer nodes (labelled as H1 through H8), which was 
 found to yield the best accuracy among the tested neural nets as shown in the left panel. The input features (labelled as I1--I10) 
 are the ten continuum flux densities from FFA16, and the outputs are the protostellar classes (Class~0, Class~I, and flat-spectrum sources, 
 labelled as O1, O2, and O3, respectively). The black lines between the layers represent positive weights, while the grey lines indicate 
 negative weights. The thickness of the lines is proportional to the magnitude of the weight with respect to all other weights.}
\label{figure:nnet}
\end{center}
\end{figure*}

We experimented with different numbers of neurons in the hidden layer, ranging from two to nine, where the latter number is equal to 
the number of features (ten flux densities) minus one. Using too many layers or neurons in the net can lead to overfitting, 
and hence we only employd a single-hidden-layer model. As shown in the left panel in Fig.~\ref{figure:nnet}, 
the best performance was found when there were eight neurons. 
The corresponding neural net is shown in the right panel in Fig.~\ref{figure:nnet}.

\section{Results and discussion}

\subsection{Performance metrics}

\begin{figure*}%[H]
\begin{center}
\includegraphics[width=0.47\textwidth]{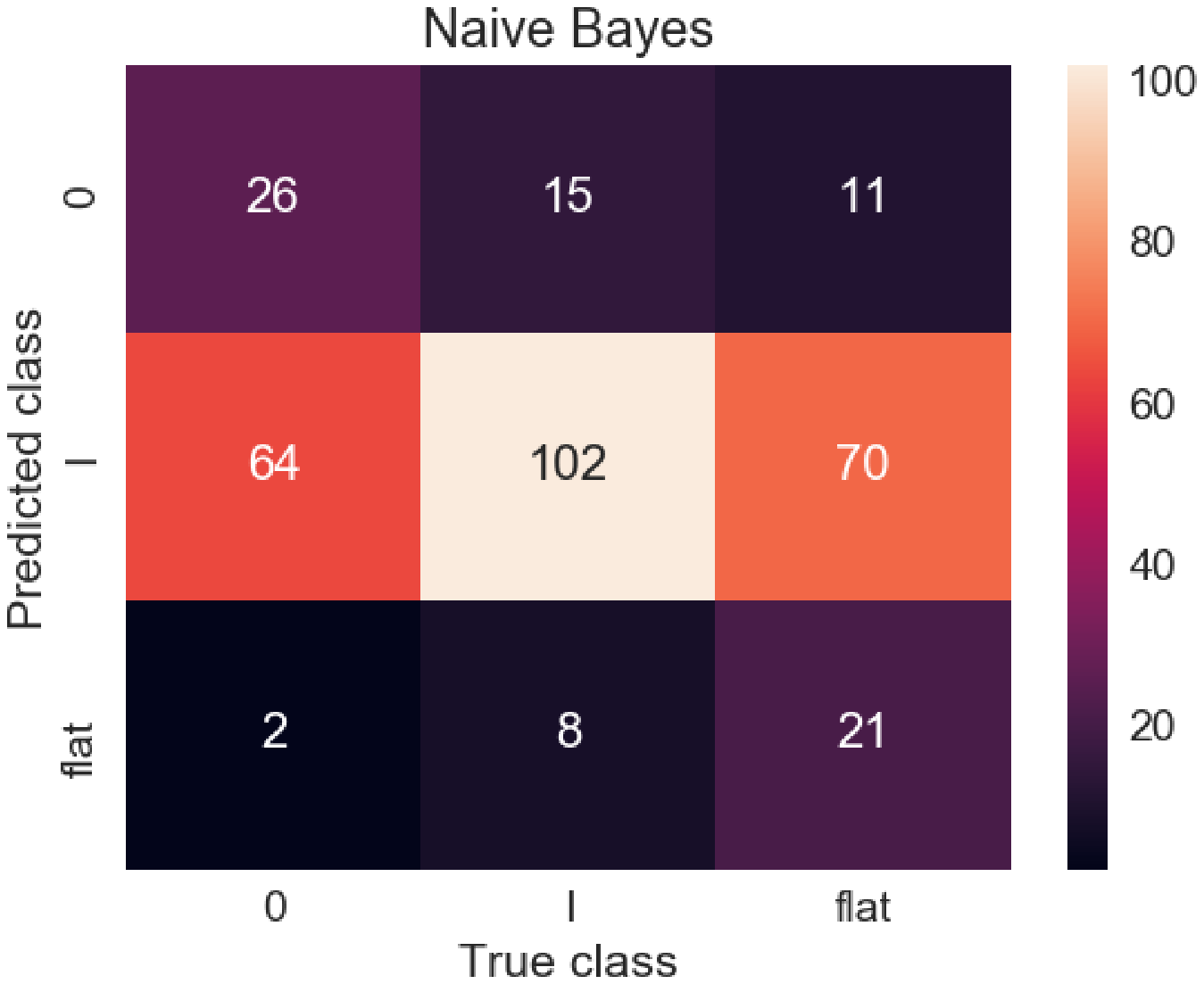}
\includegraphics[width=0.47\textwidth]{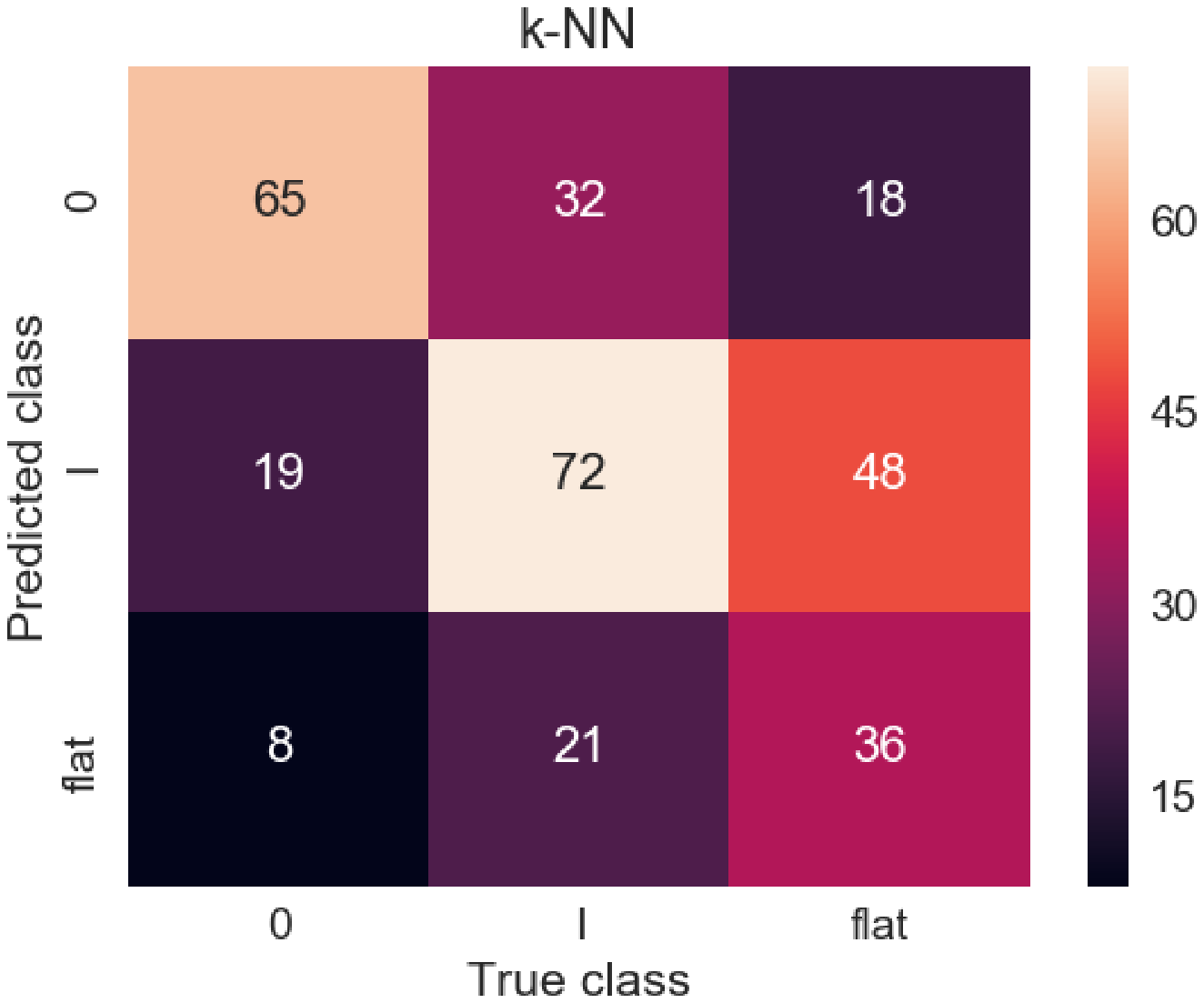}
\includegraphics[width=0.47\textwidth]{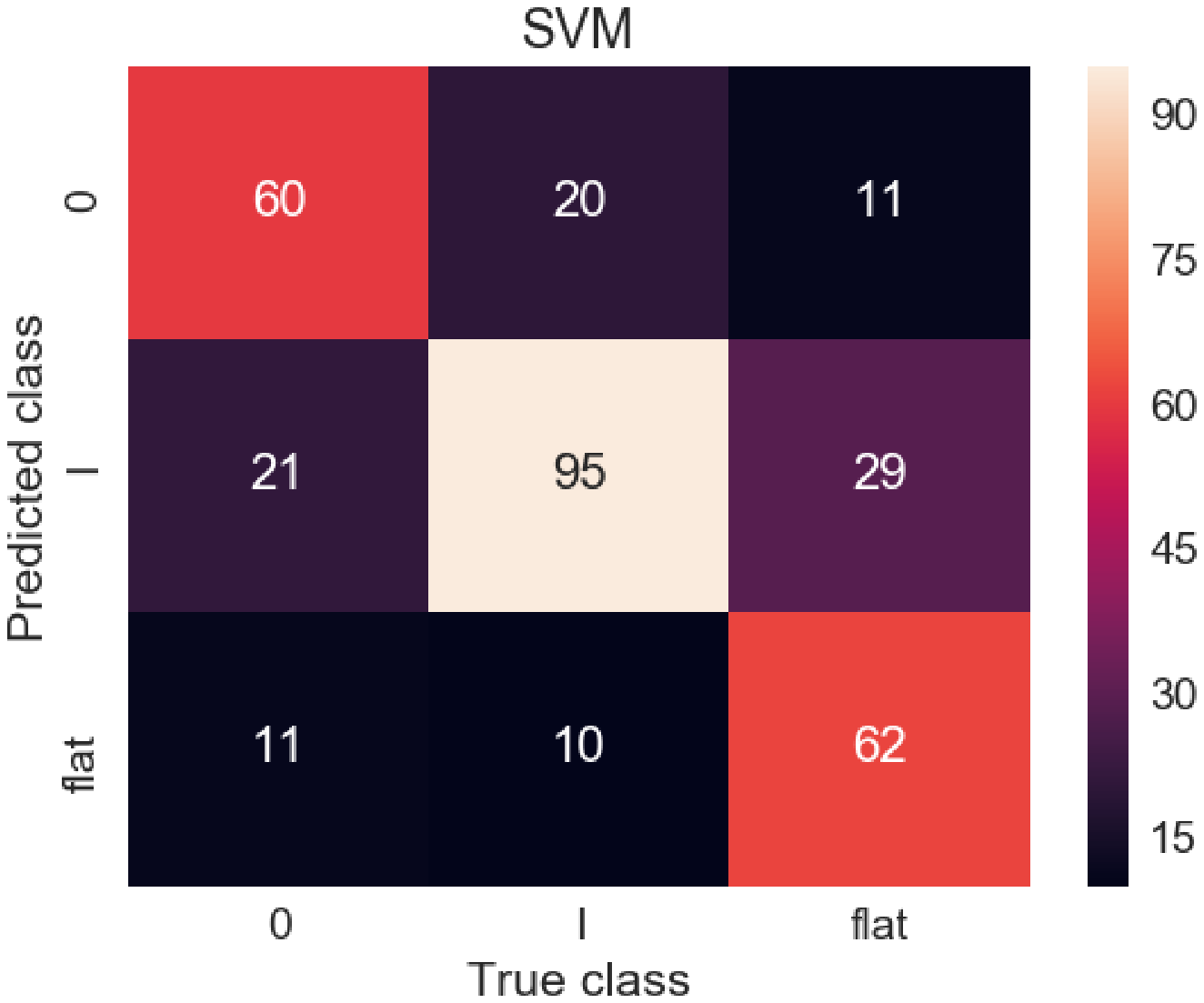}
\includegraphics[width=0.47\textwidth]{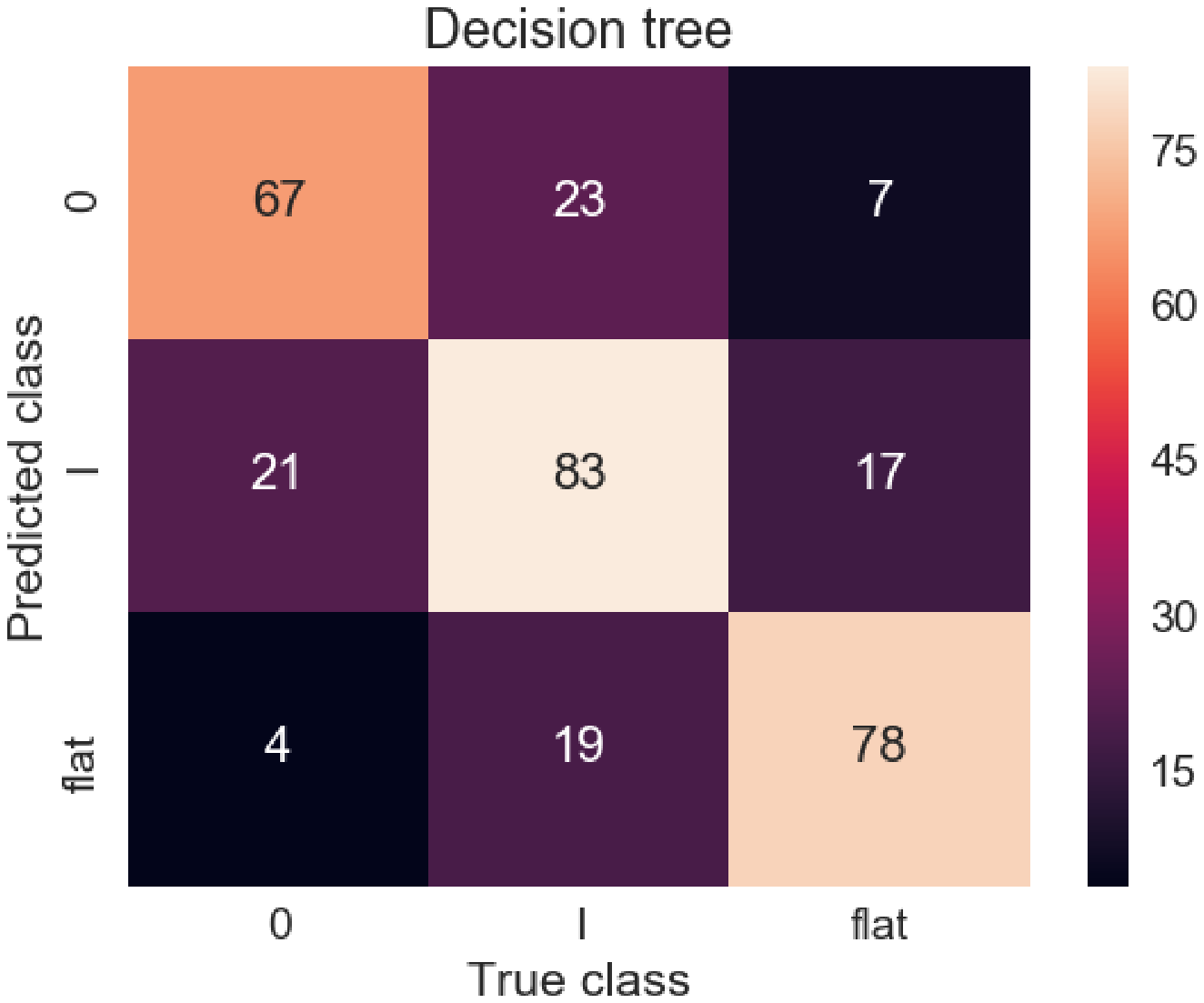}
\includegraphics[width=0.47\textwidth]{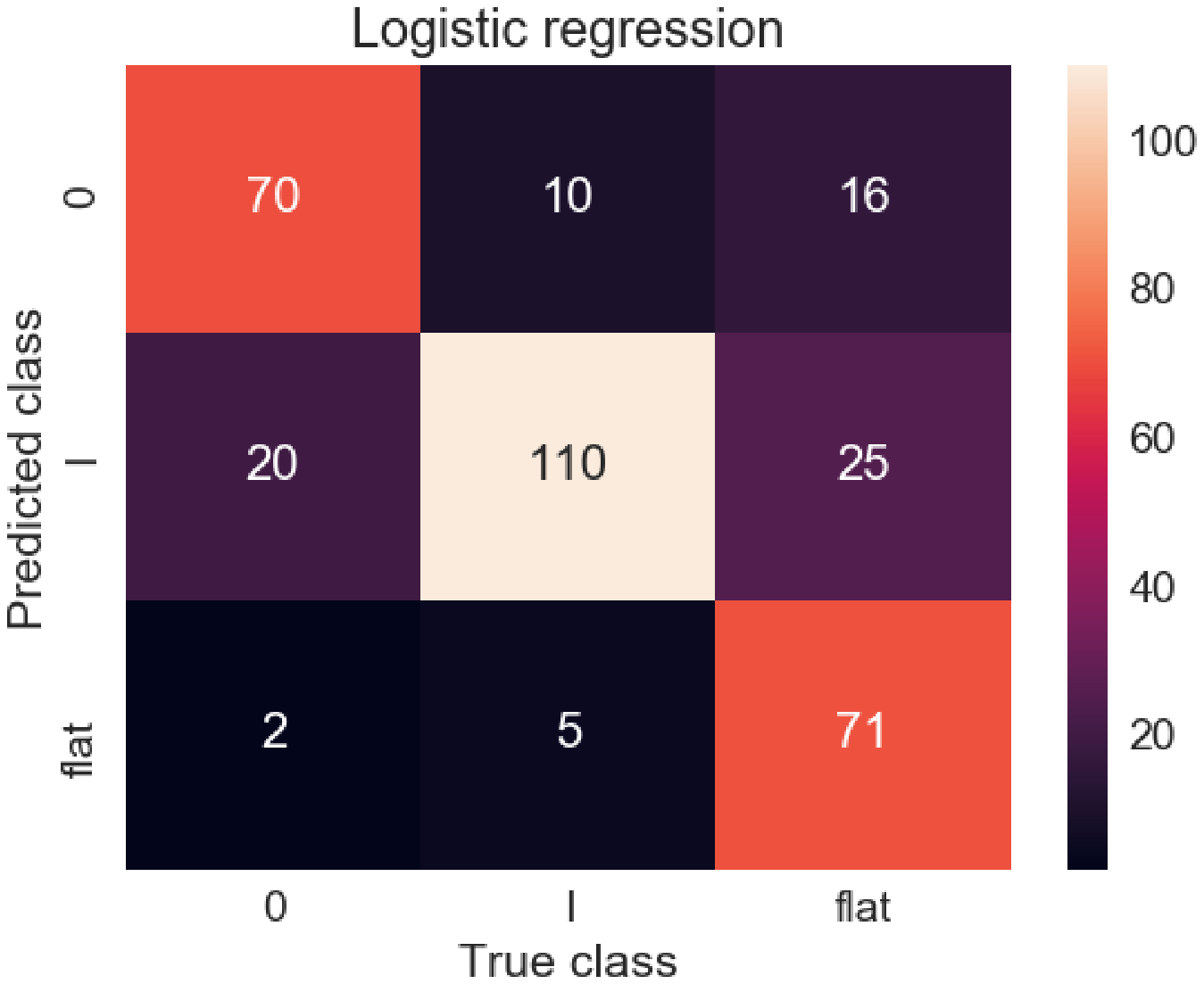}
\includegraphics[width=0.47\textwidth]{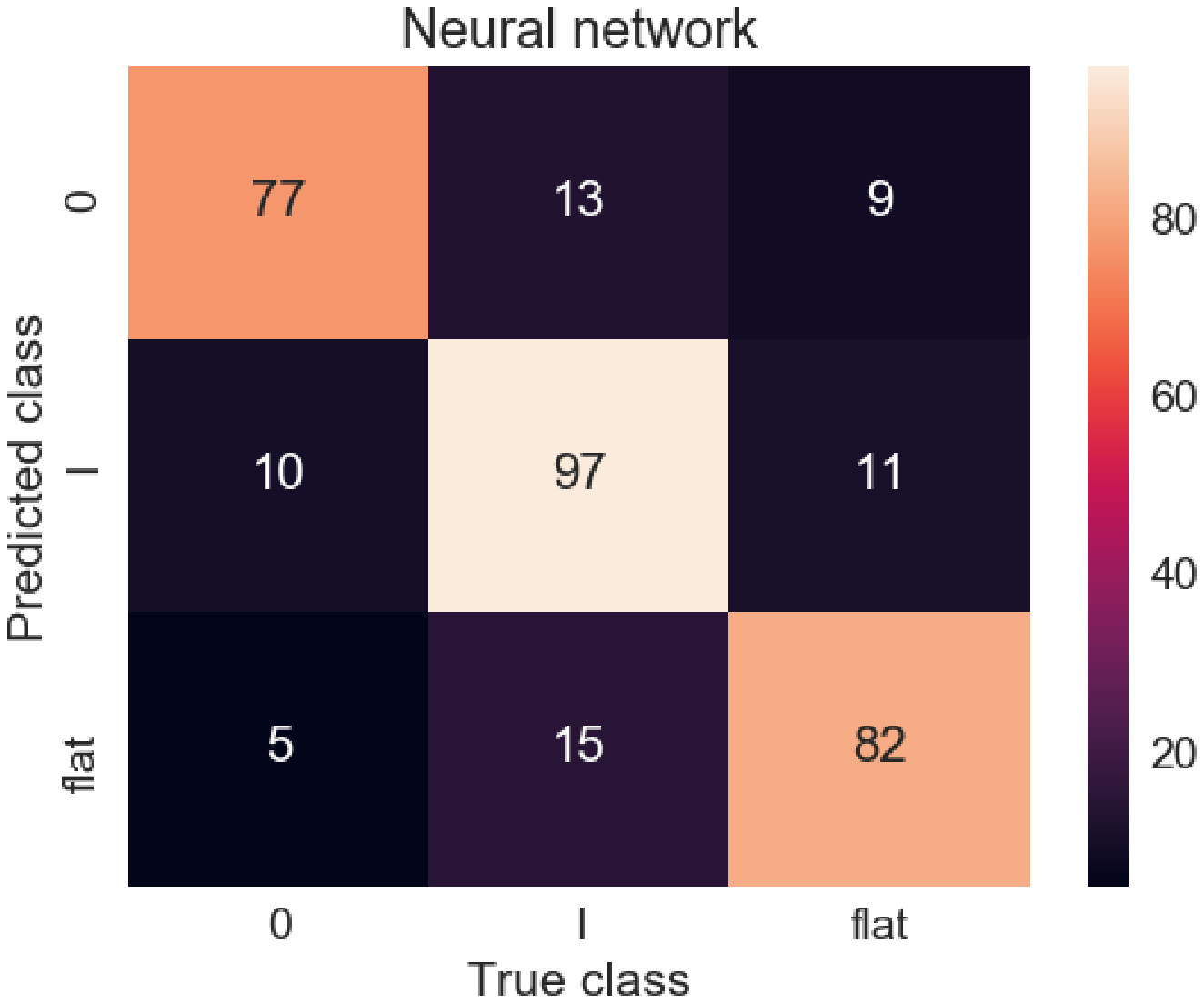}
\includegraphics[width=0.47\textwidth]{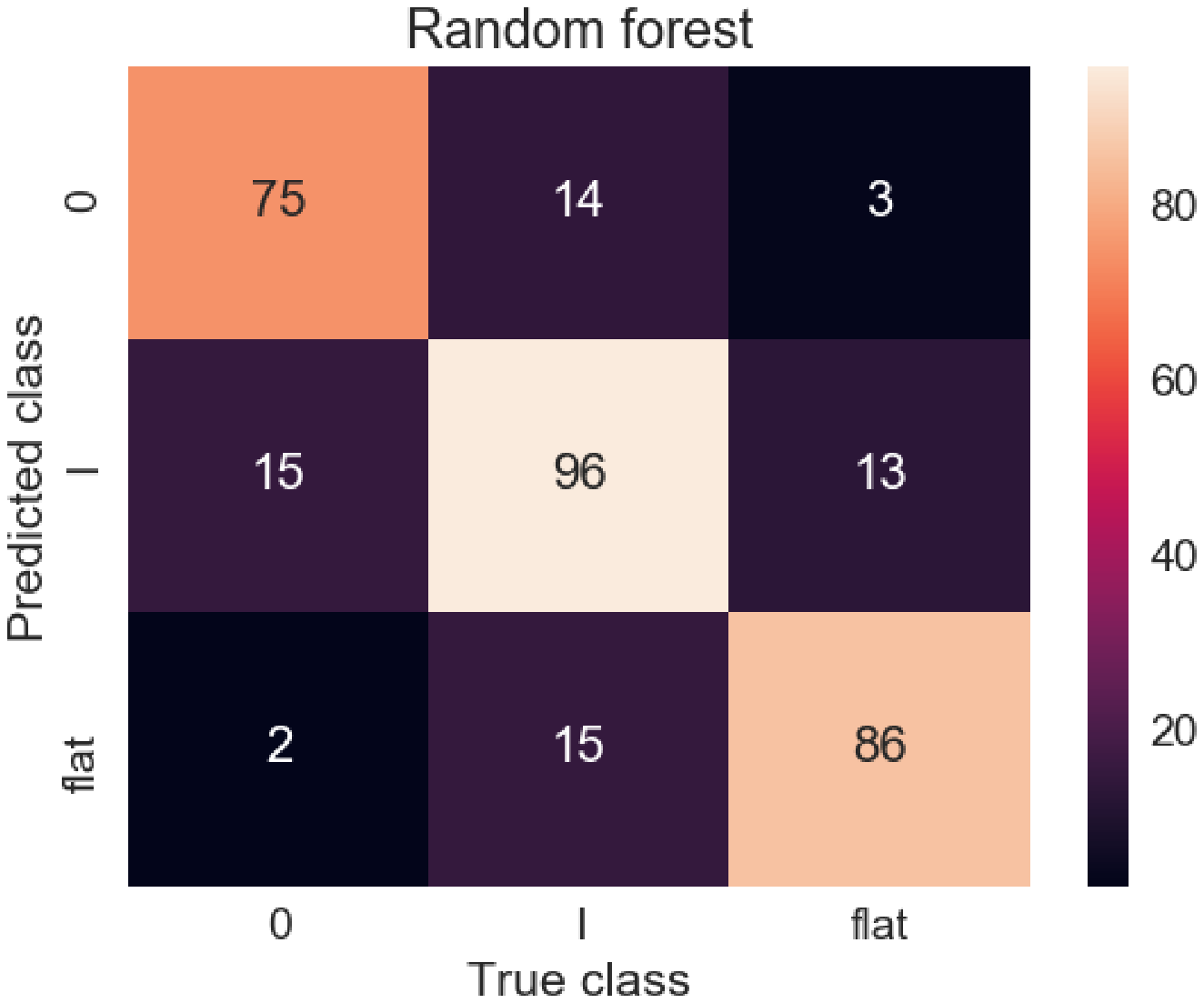}
\includegraphics[width=0.47\textwidth]{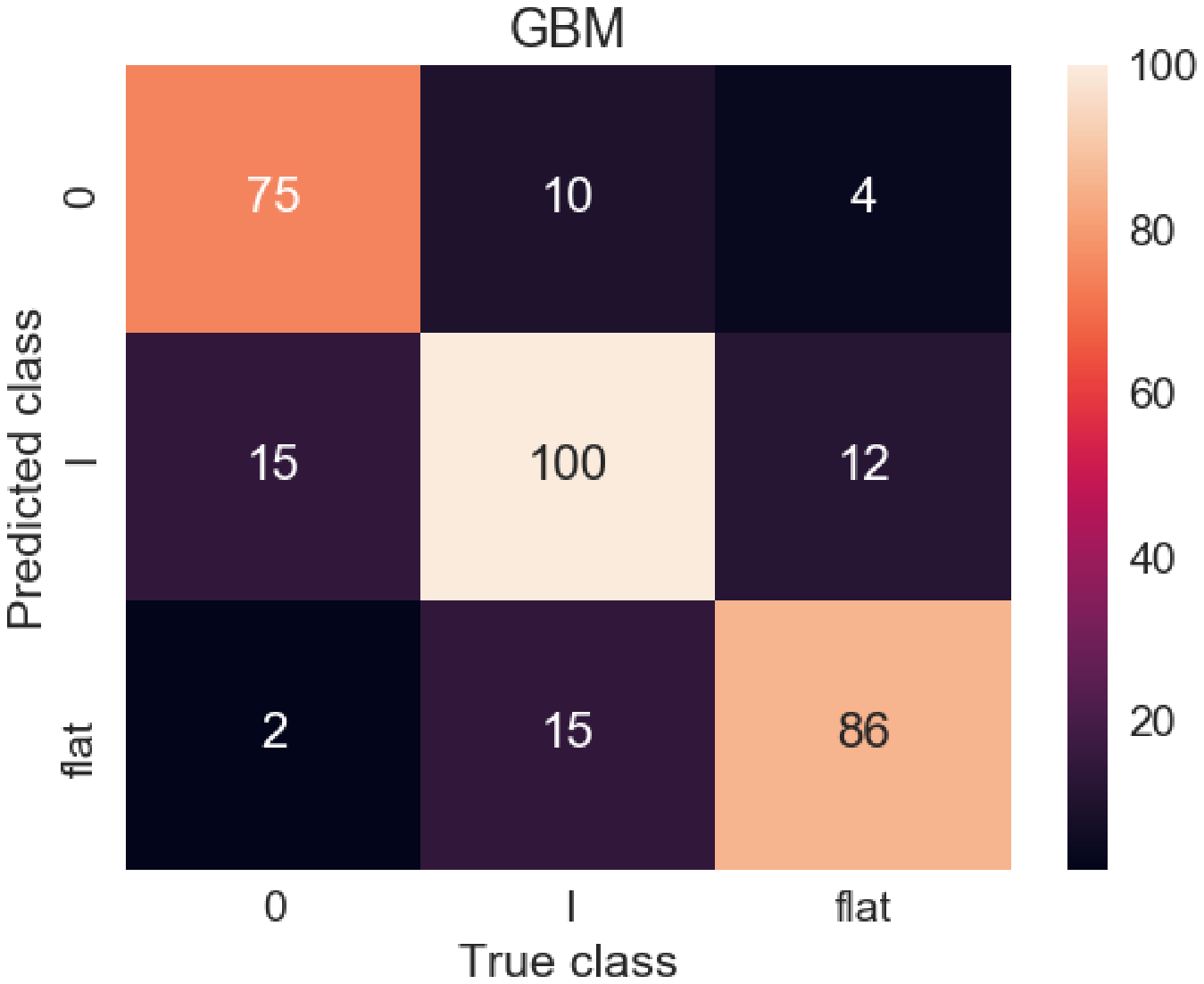}
\caption{Colour-coded confusion matrices showing the performance of the tested machine learning classifiers.}
\label{figure:cm}
\end{center}
\end{figure*}

To quantify and visualise the performance of each of the aforementioned classification algorithm, we derived their confusion matrices 
(see Fig.~\ref{figure:cm}). The columns of each matrix represent the instances in an actual, SED-based class (FFA16), 
while the rows represent the instances in a predicted class. The diagonal elements of a confusion matrix show the cases 
where the predicted class is the same as the true (SED-based) class, while the off-diagonal elements represent the misclassified 
cases. Hence, the larger the diagonal element numbers are, the better the classifier has performed. 

Several different parameters that characterise the performance of a classifier can be calculated from the confusion matrix 
(e.g. \citealp{fawcett2006}), but here we focus only on four of them, namely the aforementioned overall accuracy, 
which tells how often the classifier is correct (the sum of the diagonal elements of the confusion matrix divided by the total 
number of cases), purity of a class (ratio between the correctly classified sources of a class and the number of sources 
classified in that class), contamination of a class (ratio between the misclassified sources in a class and the number of sources 
classified in that class), which is given by ${\rm contamination}=1-{\rm purity}$ (e.g. \citealp{tangaro2015}), 
and the Matthews correlation coefficient (\citealp{matthews1975}), or the phi coefficient, which is defined as

\footnotesize
\begin{equation}
\label{eq:mcc}
{\rm MCC}=\frac{{\rm TP \times TN - FP \times FN}}{\sqrt{{\rm (TP+FP)\times(TP+FN)\times(TN+FP)\times(TN+FN)}}}\,,
\end{equation}
\normalsize
where TP, TN, FP, and FN are the numbers of true positives, true negatives, false positives, and false negatives, respectively.
The MCC can be considered a correlation coefficient between the true and predicted binary classifications, and its value lies 
in the range of $-1$ to $+1$, where $-1$ means a full disagreement between the predicted and true classes, 0 is equivalent to 
random guess, and $+1$ indicates a perfect prediction performance. Because we are dealing with a multiclass classification 
(rather than binary classification), we calculated the so-called  micro-averaged MCCs, that is we summed all the TP, TN, FP, and FN values for each class to calculate the MCC. More precisely, the TPs were derived by taking the sum of the confusion matrix diagonal elements, the TNs were calculated by removing the target class' row and column from the confusion matrix, and then taking the sum of all the remaining elements, 
the FPs were calculated as the sum of the respective column, minus the diagonal element for the class under consideration, 
and the FNs were computed by taking the sum of the respective row elements, minus the diagonal elements. 
The performance metrics are tabulated in Table~\ref{table:acc}. 

\subsection{Performance of the tested classifiers}

The evolutionary stages of our target protostellar objects were originally derived by FFA16 using an SED analysis. 
Hence, by comparing our classifications with respect to these SED-based classes, we are assuming that the SED classes 
are correct. Although panchromatic SEDs are expected to yield some of the most reliable source evolutionary stages 
(if not even the most reliable ones), it is still good to keep in mind the possibility that a machine learning 
classifier could predict a correct evolutionary stage for a source although it would differ from its SED class. The SED-based 
source classification itself can depend on the exact method of how the analysis is performed (e.g. modified blackbody fitting 
versus fitting based on radiative transfer models as done in FFA16). Moreover, the assumptions about the dust grain properties 
affect the dust-based physical properties of the source, and hence the corresponding evolutionary stage. 
From an observational point of view, the source inclination angle and variability might also affect 
the inferred evolutionary stage (e.g. the central protostar might be visible if observed through the outflow cavity). Related to the issues of the SED analysis, we remind the reader that all the FFA16 sources were assumed to lie at the 
same distance (Sect.~2). Hence, we did not use the distance as a separate feature in our supervised source classification. However, 
if the source sample is being drawn from different star-forming regions that lie at different distances from the Sun, the distance should 
be included as a feature because some of the fundamental source properties depend on it (e.g. the mass scales as $d^2$).

In the following subsections, we briefly discuss the performance of each tested classifier. 
The algorithms are discussed in the order of increasing classification accuracy, which were found to range 
from 47\% to 82\% (Table~\ref{table:acc}).

\subsubsection{Na\"ive Bayes}

The poorest job in the present classification analysis was done by the na\"ive Bayes classifier with an accuracy of only 47\% and an MCC of 0.20. For comparison, among three possible classes as in the present study (Class~0, Class~I, and flat sources), 
the accuracy of random guess would be $\sim33\%$. As described in Sect.~3.5.5, na\"ive Bayes classifier is based on the assumption 
that the predictors are independent of each other. Considering the present set of features, which is composed of continuum flux densities, the assumption of their independence is certainly violated. 
For example, as mentioned in Sect.~3.2, the far-IR and submm flux densities explored in the present work depend on each other via dust opacity.

\subsubsection{$k$-nearest neighbours}

With an accuracy of only 54\% and an MCC of 0.31, the $k$-NN classifier was found to be comparable 
to the na\"ive Bayes classifier. As described in Sect.~3.5.6, the number of nearest neighbours we considered was set to $k=7$.
The decision of how many neighbours to take into account controls the model's ability to generalise to future data instances. 
Although the exact value of the optimal $k$ is dependent on the analysed data set, there are a few general things to keep in mind. First, 
if only a single nearest neighbour is considered, the classifier is subject to noisy data features. For this reason, we did not 
adopt the value $k=1$ although it yielded a better accuracy than using $k=7$. Secondly, 
while a large $k$ reduces the influence of noisy data, it is computationally more expensive and 
suffers from the possibility of ignoring important, small-scale patterns (and one might end up considering cases that are not even actual neighbours anymore). 
Hence, the optimal $k$ is expected to lie somewhere between these two extreme cases (e.g. \citealp{lantz2015}). A common 
empirical rule of thumb is to set $k$ equal to $\sqrt{n_{\rm train}}$, where $n_{\rm train}$ is the size of 
the training data set (e.g. \citealp{hassanat2014}, and references therein; \citealp{lantz2015}). 
This will usually lead to large values of $k$ (i.e. many neighbours), 
which reduces the effect of variance caused by noisy data. In our 10-fold CV analysis, 
nine folds were used to train the $k$-NN classifier, which means that $n_{\rm train}$ was roughly $\sim270$, 
which would suggest a value of $k\simeq16$. The optimal number of nearest neighbours we found (in terms of accuracy), 
$k=7$, is over two times smaller than suggested by the aforementioned rule of thumb. Hence, although possibly being time-consuming, 
the best value of $k$ should probably be searched using a cross-validation approach as in the present study 
(see Fig.~\ref{figure:acc_knn}). 

\subsubsection{Support vector machine}

The accuracy of our SVM classifier, 68\%, is a factor of 1.26 better than that of our $k$-NN classifier. 
Moreover, the MCC of our SVM ($\textrm{MCC}=0.52$) just exceeds a binary classification threshold of 0.50 
between pure guessing ($\textrm{MCC=0}$) and perfect prediction ($\textrm{MCC=1}$). By tuning the hyperparameters of the SVM, 
such as the $C$ parameter, the classification accuracy could potentially be improved, although the risk for 
overfitting might increase accordingly.

\subsubsection{Decision tree}

A simple decision tree algorithm was found to perform fairly well as compared to the other algorithms tested 
in the present work. The accuracy and MCC (0.71 and 0.57) of our decision tree classifier are comparable to those derived 
for the SVM.

\subsubsection{Logistic regression}

Our multinomial logistic regression yielded a classifcation accuracy of 79\% with fairly pure classes 
(purity is 0.81, 0.71, and 0.91 for the Class~0, Class~I, and flat-spectrum sources, respectively). The derived 
MCC of 0.68 is a factor of 1.19 larger than for our decision tree model.

\subsubsection{Neural network}

Eighty percent of the test cases were correctly classified by our neural network classifier. Also, an MCC of 0.70 
derived for the classifier shows that its prediction performance is fairly good among the tested algorithms. Overall, the 
performance metrics of our neural network are comparable to our logistic regression. This is perhaps unsurprising, 
because the transfer function in our neural net was taken to be the logistic sigmoid function.

\subsubsection{Random forest}

The second highest classification accuracy (81\%) and an MCC (0.71) were derived for our random forest classifier. 
As expected, a random forest classifier did a much better job than a simple decision tree, which is an indication that the 
former generalises better on unseen data than the latter.

\subsubsection{Gradient boosting}

The best classification performance (82\% accuracy) was obtained with a GBM.
Our gradient boosting classifier led to fairly pure classifications per class (0.84, 0.79, and 0.83 for the Class~0, Class~I, 
and flat-spectrum sources), and hence low contaminations. Its MCC of 0.73 also indicates a reasonable prediction performance. 
As expected, GBM outperforms the simple decision tree (a factor of 1.15 better accuracy), while our GBM was found to be only 
marginally (factor of 1.01) more accurate than our random forest classifier with an accuracy of 81\% and MCC of 0.71 (the second most accurate 
classifier in the present study).

\begin{table*}
\caption{Performance metrics of the tested machine learning classifiers in order of increasing accuracy.}
{\normalsize
\centering
\label{table:acc}
\begin{tabular}{c c c c c}
\hline\hline 
Classifier & Accuracy & Purity\tablenotemark{a} & Contamination\tablenotemark{a} & MCC\\  
\hline
Na\"ive Bayes & 0.47 & 0.50, 0.69, 0.68 & 0.50, 0.53, 0.32 & 0.20 \\ [1ex]
$k$-NN & 0.54\tablenotemark{b} & 0.57, 0.52, 0.55 & 0.43, 0.48, 0.45 & 0.31\\ [1ex]
SVM & 0.68 & 0.66, 0.66, 0.75 & 0.34, 0.34, 0.25 & 0.52 \\ [1ex]
Decision tree & 0.71 & 0.69, 0.17, 0.77 & 0.31, 0.83, 0.23 & 0.57 \\[1ex]
Logistic regression & 0.79 & 0.81, 0.71, 0.91 & 0.19, 0.29, 0.09 & 0.68 \\[1ex]
Neural network & 0.80\tablenotemark{c} & 0.78, 0.82, 0.80 & 0.22, 0.18, 0.20 & 0.70\\[1ex]
Random forest & 0.81 & 0.82, 0.77, 0.85 & 0.18, 0.23, 0.15 & 0.71 \\[1ex]
GBM & 0.82 & 0.84, 0.79, 0.83 & 0.16, 0.21, 0.17 & 0.73 \\[1ex]
\hline
\end{tabular} 
\tablenotetext{a}{The three values reported for the purity and contamination refer to the Class~0, Class~I, and flat-spectrum sources.}\tablenotetext{b}{This accuracy was reached using a value of $k=7$ (Fig.~\ref{figure:acc_knn}). With $k=1$, the accuracy was about 0.55, but such model is likely subject to overfitting (Sect.~3.5.6).}\tablenotetext{c}{This accuracy was reached with eight neurons in a single-hidden-layer neural net (see Fig.~\ref{figure:nnet}).}
}
\end{table*}
 
\subsection{Unveiling the most important wavelength bands for classifying young stellar objects via supervised learning algorithms}

The PCA presented in Sect.~3.4 suggests that the 3.6~$\mu$m flux density is the most informative feature in the preset study 
(the feature explains 56.1\% of the variance of the data; see Fig.~\ref{figure:pca}). For comparison, 
the next most important band was found to be the 4.5~$\mu$m band, which explains 18.1\% of the data variance. 
As an alternative approach to unveil the most important band for the present YSO classification, 
we employed the leave-one-out cross-validation (LOOCV) technique, that is the classifications 
were done by using nine out of the ten features (ranging from the \textit{Spitzer}/IRAC 3.6~$\mu$m band to LABOCA 870~$\mu$m band), 
and this process was repeated ten times with a different wavelength band left out every time.

\begin{figure*}%[H]
\begin{center}
\includegraphics[width=0.9\textwidth]{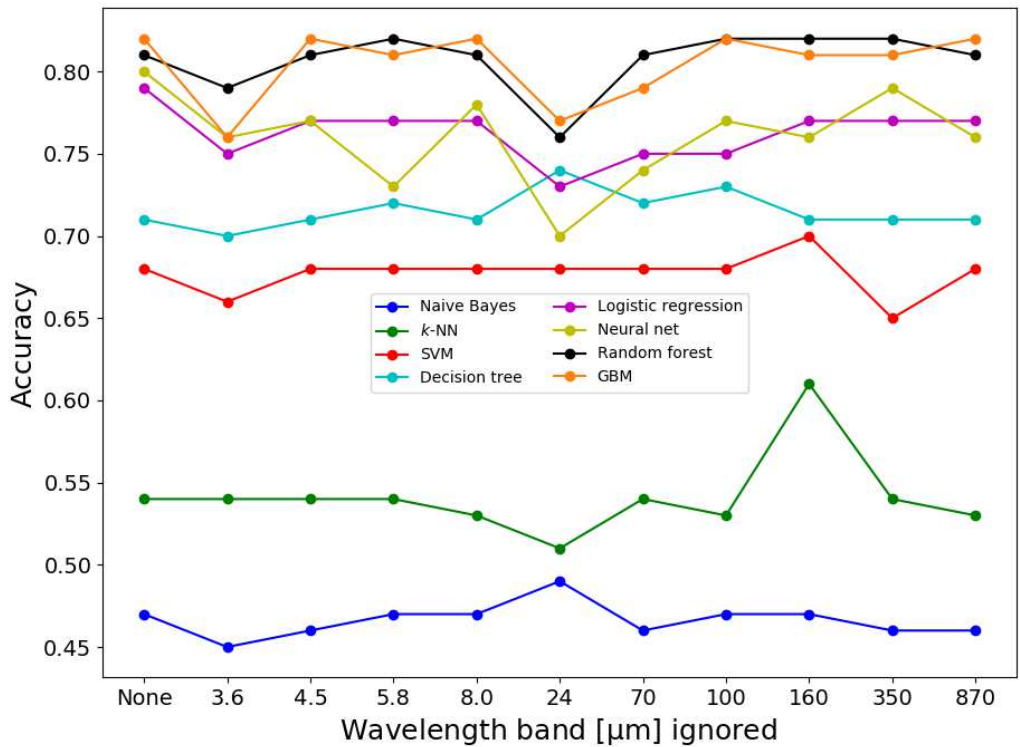}
\caption{Behaviour of the classification accuracy when one of the ten continuum bands was left out of consideration. For reference, 
the first data point from left shows the accuracy when all the ten features were used (see column~(2) in Table~\ref{table:acc}).}
\label{figure:loocv}
\end{center}
\end{figure*}

In Table~\ref{table:loocv}, we tabulate the classification accuracies of each of the tested algorithm when one out of the ten most relevant features was being left out. The results are presented visually in Fig.~\ref{figure:loocv}.
We note that in the cases of the $k$-NN and neural network classifiers, the hyperparameter settings (the $k$ value and number of neurons) 
were identical to those used for the full feature set. 
The results suggest that if the \textit{Spitzer}/IRAC 3.6~$\mu$m band 
is ignored, the classification accuracies drop with respect to the full feature set (though only by factors of 1--1.08), which conforms with the PCA result of 3.6~$\mu$m band 
being the most informative one. The importance of 
the 3.6~$\mu$m band could, at least partly, be related to the shocked H$_2$ emission associated with protostellar outflows, 
although such shock emission is stronger at 4.5~$\mu$m (e.g. \citealp{ybarra2009}). There is also a polycyclic aromatic 
hydrocarbon feature at 3.3~$\mu$m owing to C-H stretching that might contribute to the \textit{Spitzer}/IRAC band 1 emission 
(e.g. \citealp{draine2003} for a review). Also, the \textit{Spitzer}/MIPS 24~$\mu$m band, which is sensitive to warm dust emission 
(e.g. Rathborne et al. 2010), appears 
to be a fairly important feature for most of the classifiers; if the band is ignored, the classification accuracies 
drop by factors of 0.96--1.14. For the decision tree and na\"ive Bayes classifiers, however, the classification accuracy 
was actually marginally higher when the 24~$\mu$m data were ignored (by a factor of 1.04 in both cases).
From a physical point of view, the most important 
wavelength bands are expected to be those probing the peak of the source SED (i.e. around 
$\sim100$~$\mu$m; see Sect.~2), but this is not manifested in our LOOCV feature selection.

As mentioned above, in some cases it was found that the exclusion of a wavelength band actually 
increases the classification accuracy with respect to the case where all the ten 
features are being used. Most notably, this happens when the \textit{Herschel}/PACS 
160~$\mu$m band is ignored from the $k$-NN classification (the accuracy increases 
by a factor of 1.13; see the green curve in Fig.~\ref{figure:loocv}). This suggests that the inclusion 
of the aforementioned band might have led to a slight overfitting effect in our $k$-NN classification. 
We also note that the random forest and GBM are generally found to yield the best classification accuracies when ignoring the different 
features, but in the case where the 3.6~$\mu$m band was ignored, our neural network appeared equally good as the GBM (76\% accuracy for both). 
In the latter case, the random forest was found to yield the highest classification accuracy of 79\%, 
but the difference is only marginal with respect to the GBM. Also, these small differences in the accuracies compared to the 
usage of all the ten features can not be considered significant owing to the random sampling nature of the both the random forest algorithm 
and the 10-fold CV technique.

Of course, there are many feature combinations among the considered flux densities that could be explored. 
For example, there are 45 unique flux density pairs in the present set of ten different flux densities. However, 
a thorough feature analysis is beyond the scope of the present study.

Considering the future YSO surveys where similar machine learning approaches could be used as in the present study, 
the observed wavelengths are likely to differ from those we have analysed (e.g. the cryogenic phase of \textit{Spitzer} 
operated from 2003 to 2009, while its warm mission (using the 3.6~$\mu$m and 4.5~$\mu$m IRAC bands only) is scheduled to end 
in March 2019 (e.g. \citealp{hora2012}; \citealp{yee2017}), and the \textit{Herschel} mission ended on 29 April 2013 when the satellite 
ran out of its helium coolant (e.g. \citealp{sauvage2014})). Nevertheless, the aforementioned analysis suggests 
that bands near 3.6~$\mu$m and 24~$\mu$m would be informative if the other bands are comparable to those employed here.

\begin{table*}
\caption{Classification accuracy when one of the ten continuum bands was left out of consideration.}
{\normalsize
\centering
\label{table:loocv}
\begin{tabular}{c c c c c c c c c c c}
\hline\hline 
Classifier & \multicolumn{10}{c}{Wavelength band [$\mu$m] ignored} \\
           & 3.6\tablenotemark{a} & 4.5\tablenotemark{a} & 5.8\tablenotemark{a} & 8.0\tablenotemark{a} & 24\tablenotemark{b} & 70\tablenotemark{c} & 100\tablenotemark{c} & 160\tablenotemark{c} & 350\tablenotemark{d} & 870\tablenotemark{d} \\
\hline
Na\"ive Bayes & 0.45 & 0.46 & 0.47 & 0.47 & 0.49 & 0.46 & 0.47 & 0.47 & 0.46 & 0.46 \\ [1ex]
$k$-NN\tablenotemark{e} & 0.54 & 0.54 & 0.54 & 0.53 & 0.51 & 0.54 & 0.53 & 0.61 & 0.54 & 0.53\\ [1ex]
SVM & 0.66 & 0.68 & 0.68 & 0.68 & 0.68 & 0.68 & 0.68 & 0.70 & 0.65 & 0.68 \\ [1ex]
Decision tree & 0.70 & 0.71 & 0.72 & 0.71 & 0.74 & 0.72 & 0.73 & 0.71 & 0.71 & 0.71\\[1ex]
Logistic regression & 0.75 & 0.77 & 0.77 & 0.77 & 0.73 & 0.75 & 0.75 & 0.77 & 0.77 & 0.77\\[1ex]
Neural network\tablenotemark{f} & 0.76 & 0.77 & 0.73 & 0.78 & 0.70 & 0.74 & 0.77 & 0.76 & 0.79 & 0.76\\[1ex]
Random forest & 0.79 & 0.81 & 0.82 & 0.81 & 0.76 & 0.81 & 0.82 & 0.82 & 0.82 & 0.81\\[1ex]
GBM & 0.76 & 0.82 & 0.81 & 0.82 & 0.77 & 0.79 & 0.82 & 0.81 & 0.81 & 0.82 \\[1ex]
\hline
\end{tabular} 
\tablenotetext{a}{\textit{Spitzer}/IRAC band.}\tablenotetext{b}{\textit{Spitzer}/MIPS band.}\tablenotetext{c}{\textit{Herschel}/PACS band.}\tablenotetext{d}{APEX bolometer band.}\tablenotetext{e}{The number of nearest neighbours was fixed at $k=7$.}\tablenotetext{f}{The number of nodes in the hidden layer was fixed at eight.}
}
\end{table*}

\subsection{Potential of using machine learning in classifying young stellar objects}

Owing to the fact that we considered only ten out of the original 29 features (i.e. the 2MASS and \textit{Spitzer}/IRS flux 
densities were ignored), a protostellar classification accuracy of 82\% with a GBM can be considered fairly good. 
Also, although being the largest, homogeneous protostar sample drawn from a single molecular cloud system, 
the size of the employed data set is fairly small for a machine learning approach (only 319 sources in total), and 
undoubtedly a higher classification accuracy could be reached with a larger training sample. On the other hand, as shown in 
Fig.~\ref{figure:scatter}, the Orion Class~0, Class~I, and flat-spectrum sources from FFA16 are not well 
separable in the two-dimensional projections of the feature space, which is an indication that learning their 
classification is demanding for at least some of the supervised classifiers.

A more detailed missing value imputation, such as estimating the missing far-IR to 
submm flux densities in a source-by-source fashion 
from the existing values by assuming a value of $\beta$ (see Sect.~3.2), could lead to an improved performance. However, this approach is 
also based on assumptions about the flux frequency dependence at different bands. Finally, an application of more advanced 
ensemble methods, like the extreme gradient boosting (XGBoost; \citealp{chen2016}), has the potential to lead to an improved accuracy. 

As mentioned in Sect.~4.3, the future YSO surveys where machine learning assisted source classification could be used will 
undoubtedly be carried out at least partly at different wavelengths than considered here. This also means that the models trained 
in the present work can not be employed as such, but they would need to be retrained (cf.~the deep learning knowledge transfer study by Dom{\'{\i}}nguez 
S{\'a}nchez et al. (2018)). In fact, even in a hypothetical case where there would 
be a survey of YSOs carried out at exactly the same ten bands as we considered, the rms noise levels of the observations would probably 
still be different, which would again call for retraining the classifiers (for example, deeper surveys could detected weaker sources than those 
in the FFA16 Orion sample). Regarding this issue, we note that the $10\sigma$ limiting magnitudes in the \textit{Spitzer}/IRAC 3.6, 4.5, 
5.8, and 8~$\mu$m data, and the \textit{Spitzer}/MIPS 24~$\mu$m data employed by FFA16 were 16.5, 16.0, 14.0, 13.0, and 8.5~mag, 
respectively (\citealp{megeath2012}; \citealp{spezzi2015}), and Megeath et al. (2012) derived the final magnitudes for all their 
\textit{Spitzer} sources that were detected with uncertainties of $\leq 0.25$~mag in one of the four IRAC bands. The properties of the 
\textit{Herschel} and APEX data products employed by FFA16 will be described in more detail by B.~Ali et al. (in prep.) and 
T.~Stanke et al. (in prep.); see Fischer et al. (2017). Moreover, the source flux densities depend on the aperture sizes used 
to extract the photometry, and if the apertures differ from those used by FFA16 (who, for example, used the aperture radii of $9\farcs6$, $9\farcs6$, 
and $12\farcs8$ for the \textit{Herschel} 70, 100, and 160~$\mu$m data, respectively), the present classifiers would again have 
to be retrained.

Although reaching high accuracies, a supervised machine learning classification cannot replace an SED-based YSO classification
because an SED analysis also yields the important physical properties of the source, like the dust temperature and 
(envelope) dust mass (however, the SED fitting itself could also rely on machine learning regression). Moreover, SED analyses are still 
expected to be needed to create the training data sets for machine learning applications. 
Nevertheless, if appropriate training data sets are available, machine learning techniques can serve as a quick way to estimate the relative percentages of YSOs in different 
evolutionary stages in the era of big astronomical data (e.g. \citealp{hocking2018}; \citealp{pearson2018}), 
and also to mine the YSO data to unveil interesting subsamples for more detailed follow-up observations (e.g. \citealp{marton2016}).  

\section{Summary and conclusions}

We used eight different supervised machine learning algorithms to classify the Orion protostellar objects from FFA16 into 
Class~0, Class~I, and flat-spectrum sources. On the basis of PCA, we employed only the IR and submm continuum photometric data 
from FFA16. The training and testing of the classifiers were performed by using a 10-fold CV technique.
Using the SED-based classifications of FFA16 as the benchmark, we found that the highest classification accuracy is reached by 
a GBM algorithm (82\% of the cases were correctly classified with $\gtrsim80\%$ purity and an MCC of 0.73), 
while the poorest performance was that of na\"ive Bayes classification 
(47\% accuracy). 

Our analysis suggests that among the ten continuum emission bands used in the classification, 
the \textit{Spitzer} 3.6~$\mu$m and 24~$\mu$m flux densities are the most informative features in terms of the source classification 
accuracy. Hence, these two wavelength bands would be useful to include in a panchromatic YSO classification study, especially if the other 
bands available are comparable to those analysed in the present work (i.e. 4.5, 5.8, 8.0, 70, 100, 160, 350, and 870~$\mu$m).

Larger data sets, detailed missing value imputations, and more sophisticated learning algorithms 
have the potential to improve the classification accuracies. Overall, machine learning algorithms can provide a fast 
(at least compared to an SED analysis) way to classify large samples of sources into different evolutionary stages, 
and hence estimate the statistical lifetimes of the sources, and also pick up subsamples of interesting sources for 
targeted follow-up studies. However, an obvious challenge of supervised machine learning classification is the 
creation of training data sets, which requires classification of large numbers of YSOs into different evolutionary stages on the 
basis of their measured flux densities at the observed wavelengths. Because the latter is based on SED fitting, which itself 
requires knowledge of the source distance and assumptions about the underlying dust grain model, an SED analysis 
might be a prerequisite to the usage of machine learning classifiers on new survey data sets.

%% Acknowledgements
%
\acknowledgments
% <Acnowledgments text>

I would like to thank the referee for providing constructive comments and suggestions that helped to improve the quality of this paper.
This research has made use of NASA's Astrophysics Data System and the NASA/IPAC Infrared Science Archive, which is operated by 
the JPL, California Institute of Technology, under contract with the NASA.

%% References
%% Please cite all reference entries in the article text using \cite or
%% equivalent command. 

%%%  Using BibTeX  (Name-Year style)
%
% \bibliographystyle{spr-mp-nameyear-cnd}  %% BibTeX style
\bibliographystyle{plainnat}

% \bibliography{<bib data>}                %% BibTeX data

\begin{thebibliography}{}

\bibitem[Abdi \& Williams 2010]{abdi2010} Abdi, H., \& Williams, L.~J.\ 2010, Wiley Interdisciplinary
Reviews: Computational Statistics 2 (4), 433-459

\bibitem[Adams et al. 1987]{adams1987} Adams, F.~C., Lada, C.~J., \& Shu, F.~H.\ 1987, \apj, 312, 788 

\bibitem[Alpaydin 2010]{alpaydin2010} Alpaydin, E. 2010, \textit{Introduction to Machine Learning}, 2nd ed., The MIT Press 

\bibitem[Altman 1992]{altman1992} Altman, N.~S.\ 1992, The American Statistician, 46 (3): 175–185 

\bibitem[An et al. 2018]{an2018} An, F.~X., Stach, S.~M., Smail, I., et al.\ 2018, \apj, 862, 101

\bibitem[Andr\'e \& Montmerle 1994]{andre1994} Andr\'e, P., \& Montmerle, T.\ 1994, \apj, 420, 837 

\bibitem[Andr\'e et al. 1993]{andre1993} Andr\'e, P., Ward-Thompson, D., \& Barsony, M.\ 1993, \apj, 406, 122 

\bibitem[Andr\'e et al. 2000]{andre2000} Andr\'e, P., Ward-Thompson, D., \& Barsony, M.\ 2000, in \textit{Protostars and Planets IV}, ed.~V. Mannings,
A.~P. Boss, \& S.~S. Russell (Tuscon, AZ: Univ. Arizona Press), p.~59 

\bibitem[Andr{\'e} et al. 2010]{andre2010} Andr{\'e}, P., Men'shchikov, A., Bontemps, S., et al.\ 2010, \aap, 518, L102 

\bibitem[Aniyan \& Thorat 2017]{aniyan2017} Aniyan, A.~K., \& Thorat, K.\ 2017, \apjs, 230, 20 

\bibitem[Ball \& Brunner 2010]{ball2010} Ball, N.~M., \& Brunner, R.~J.\ 2010, International Journal of Modern Physics D, 19, 1049 

\bibitem[Beck et al. 2018]{beck2018} Beck, M.~R., Scarlata, C., Fortson, L.~F., et al.\ 2018, \mnras, 476, 5516

\bibitem[Box \& Meyer 1986]{box1986} Box, G.~E.~P., \& Meyer, R.~D.\ 1986, Technometrics, 28 (1)

\bibitem[Breinman 1997]{breinman1997} Breinman, L.\ 1997, Technical Report 486, Statistics Department, University of California, 
Berkeley, CA 94720

\bibitem[Breinman 2001]{breinman2001} Breinman, L.\ 2001, Machine Learning, 45 (1): 5–32

\bibitem[Breinman et al. 1984]{breinman1984} Breinman, L., Friedman, J.~H., Stone, C.~J., and Olshen, R.~A. 1984, 
\textit{Classification and Regression Trees}, Taylor \& Francis 

\bibitem[Burges 1998]{burges1998} Burges, C.\ 1998, Data Min Knowl Disc 2 (2): 1–47

\bibitem[Cawley \& Talbot 2010]{cawley2010} Cawley, G.~C. \& Talbot, N.~L.~C 2010, JMLR, 2079-2107

\bibitem[Chen \& Guestrin 2016]{chen2016} Chen, T. \& Guestrin, C. 2016, {\tt arXiv:1603.02754}

\bibitem[Cristianini \& Shawe-Taylor 2000]{christianini2000} Cristianini, N., \& Shawe-Taylor, J.\ 2000, 
\textit{An introduction to support vector machines and other kernel-based learning methods}, Cambridge University
Press

\bibitem[Cortes \& Vapnik 1995]{cortes1995} Cortes, C., \& Vapnik, V.~N.\ 1995, Machine Learning, 20 (3): 273–297

\bibitem[Cover \& Hart 1967]{cover1967} Cover, T., \& Hart, P.\ 1967, IEEE transactions
on information theory, 13(1), 21-27

\bibitem[Cox 1958]{cox1958} Cox, D.~R.\ 1958, J Roy Stat Soc B. 20: 215–242

\bibitem[Domingos \& Pazzani 1996]{domingos1996} Domingos, P. \& Pazzani, M.\ 1996, in \textit{Proceedings of the Thirteenth 
International Conference on Machine Learning}, ed. L.~Saitta, pp.~105–112, San Francisco, CA: Morgan Kaufmann 

\bibitem[Dom{\'{\i}}nguez S{\'a}nchez et al.(2018)]{dominguez2018} Dom{\'{\i}}nguez S{\'a}nchez, H., Huertas-Company, M., 
Bernardi, M., et al.\ 2018, \mnras, \textit{in press}, {\tt arXiv:1807.00807} 

\bibitem[Draine 2003]{draine2003} Draine, B.~T.\ 2003, \araa, 41, 241 

\bibitem[Dunham et al. 2014]{dunham2014} Dunham, M.~M., Stutz, A.~M., Allen, L.~E., et al.\ 2014, \textit{Protostars and Planets VI},
H. Beuther, R. S. Klessen, C. P. Dullemond, and Th. Henning (eds.), University of Arizona Press, Tucson, 914 pp., p.~195 

\bibitem[Dunham et al. 2015]{dunham2015} Dunham, M.~M., Allen, L.~E., Evans, N.~J., II, et al.\ 2015, \apjs, 220, 11 

\bibitem[Evans et al. 2009]{evans2009} Evans, N.~J., II, Dunham, M.~M., J{\o}rgensen, J.~K., et al.\ 2009, \apjs, 181, 321
  
\bibitem[Fawcett 2004]{fawcett2006}Fawcett, T.\ 2006, Pattern Recognition Letters, 27 (8): 861–874
   
\bibitem[Fazio et al. 2004]{fazio2004} Fazio, G.~G., Hora, J.~L., Allen, L.~E., et al.\ 2004, \apjs, 154, 10   

\bibitem[Fischer et al. 2017]{fischer2017} Fischer, W.~J., Megeath, S.~T., Furlan, E., et al.\ 2017, \apj, 840, 69 

\bibitem[Friedman 1999]{friedman1999} Friedman, J.~H.\ 1999, Computational Statistics and Data Analysis, Vol.~38, p.~367-378

\bibitem[Friedman 2001]{friedman2001} Friedman, J.~H.\ 2001, Ann. Statist., Vol.~29, No.~5, 1189-1232
  
\bibitem[Furlan et al. 2016]{furlan2016} Furlan, E., Fischer, W.~J., Ali, B., et al.\ 2016, \apjs, 224, 5 (FFA16)

\bibitem[Greene et al. 1994]{greene1994} Greene, T.~P., Wilking, B.~A., Andr\'e, P., et al.\ 1994, \apj, 434, 614 

\bibitem[G{\"u}sten et al. 2006]{gusten2006} G{\"u}sten, R., Nyman, L.~{\AA}., Schilke, P., et al.\ 2006, \aap, 454, L13 

\bibitem[Hassanat et al. 2014]{hassanat2014} Hassanat, A.~B., Mohammad, A.~A., Altarawneh, G.~A., et al. \ 2014, IJCSIS,
Vol.~12, No.~8

\bibitem[Hawkins 2004]{hawkins2004} Hawkins, D.~M. 2004, J. Chem. Inf. Comput. Sci., 44(1): 1-12

\bibitem[He \& Ma 2013]{he2013} He, H., and Ma, Y. 2013, \textit{Imbalanced Learning: Foundations, Algorithms, and Applications }, 1st Edition, Wiley-IEEE Press

\bibitem[Ho 1995]{ho1995} Ho, T.~K.\ 1995, Proceedings of the 3rd International Conference on Document Analysis and Recognition, 
Montreal, QC, 14–16 August 1995. pp.~278–282

\bibitem[Hocking et al. 2018]{hocking2018} Hocking, A., Geach, J.~E., Sun, Y., \& Davey, N.\ 2018, \mnras, 473, 1108 

\bibitem[Hora et al. 2012]{hora2012} Hora, J.~L., Marengo, M., Park, R., et al.\ 2012, \procspie, 8442, 844239 

\bibitem[Hotelling 1933]{hotelling1933} Hotelling, H.\ 1933, Journal of educational psychology 24 (6), 417

\bibitem[Houck et al. 2004]{houck2004} Houck, J.~R., Roellig, T.~L., van Cleve, J., et al.\ 2004, \apjs, 154, 18 

\bibitem[Hui et al. 2018]{hui2018} Hui, J., Aragon, M., Cui, X., \& Flegal, J.~M.\ 2018, \mnras, 475, 4494 

\bibitem[James et al. 2017]{james2017} James, G., Witten, D., Hastie, T., and Tibshirani, R. 2017, \textit{An Introduction to 
Statistical Learning with Applications in R}, 8th printing, Springer Science$+$Business Media New York

\bibitem[Jeffrey \& Rosner 1986]{jeffrey1986} Jeffrey, W., \& Rosner, R. 1986, \apj, 310, 473

\bibitem[Jolliffe 2002]{jolliffe2002} Jolliffe, I.\ 2002, \textit{Principal component analysis}, Wiley Online Library

\bibitem[Kotsiantis 2007]{kotsiantis2007} Kotsiantis, S.~B. 2007, Informatica, 31, 249-268

\bibitem[Kotsiantis et al. 2006a]{kotsiantis2006a} Kotsiantis, S.~B., Kanellopoulos, D., and Pintelas, P.~E.\ 2006a, 
GESTS International Transactions on Computer Science and Engineering, Vol.~30

\bibitem[Kotsiantis et al. 2006b]{kotsiantis2006b} Kotsiantis, S.~B., Zaharakis, I.~D., and Pintelas, P.~E.\ 2006b, 
Artif Intell Rev, 26:159–190

\bibitem[Krakowski et al. 2016]{krakowski2016} Krakowski, T., Ma{\l}ek, K., Bilicki, M., et al.\ 2016, \aap, 596, A39 

\bibitem[Lada 1987]{lada1987} Lada, C.~J.\ 1987, in IAU Symposium, Vol.~115, \textit{Star Forming Regions}, p.~1 

\bibitem[Lada \& Wilking 1984]{lada1984} Lada, C.~J., \& Wilking, B.~A.\ 1984, \apj, 287, 610 

\bibitem[Lantz 2015]{lantz2015} Lantz, B. 2015, \textit{Machine Learning with R}, Second Edition, Packt Publishing Ltd.

\bibitem[Little 1988]{little1988} Little, R.~J.~A.\ 1988, Missing data adjustments in large surveys (with discussion), Journal of Business Economics and Statistics, 6, 287–301

\bibitem[Lochner et al. 2016]{lochner2016} Lochner, M., McEwen, J.~D., Peiris, H.~V., et al.\ 2016, \apjs, 225, 31 


\bibitem[Lukic et al. 2018]{lukic2018} Lukic, V., Br{\"u}ggen, M., Banfield, J.~K., et al.\ 2018, \mnras, 
476, 246 

\bibitem[Marton et al. 2016]{marton2016} Marton, G., T{\'o}th, L.~V., Paladini, R., et al.\ 2016, \mnras, 458, 3479 

\bibitem[Matthews 1975]{matthews1975} Matthews, B.~ W.\ 1975, Biochimica et Biophysica Acta (BBA) - Protein Structure, 
405 (2): 442–451

\bibitem[McCallum \& Nigam 1998]{mccallum1998} McCallum, A., \& Nigam, K.\ 1998, 
in \textit{AAAI-98 workshop on learning for text categorization}, Vol.~752, Issue~1, pp.~41-48

\bibitem[McCulloch \& Pitts 1943]{mcculloch1943} McCulloch, W., \& Pitts, W.~H. Jr.\ 1943, Bulletin of Mathematical Biophysics, 
5 (4): 115–133 

\bibitem[Megeath et al. 2012]{megeath2012} Megeath, S.~T., Gutermuth, R., Muzerolle, J., et al.\ 2012, \aj, 144, 192 

\bibitem[Miettinen 2016]{miettinen2016} Miettinen, O.\ 2016, \apss, 361, 248 

\bibitem[Miettinen et al. 2009]{miettinen2009} Miettinen, O., Harju, J., Haikala, L.~K., et al.\ 2009, \aap, 500, 845 

\bibitem[Mitchell 1997]{mitchell1997} Mitchell, T.\ 1997, \textit{Machine Learning}, McGraw Hill

\bibitem[Mosteller \& Tukey 1968]{mosteller1968} Mosteller, F., and Turkey J.~W.\ 1968, Data analysis, including statistics, 
in \textit{Handbook of Social Psychology}, eds. G.~Lindzey and E.~Aronson, Vol.~2, Addison-Wesley

\bibitem[Murthy 1998]{murthy1998} Murthy, S.~K.\ 1998, Data Min Knowl Disc 2:345–389

\bibitem[Myers \& Ladd 1993]{myers1993} Myers, P.~C., \& Ladd, E.~F.\ 1993, \apjl, 413, L47 

\bibitem[Pashchenko et al. 2018]{pashchenko2018} Pashchenko, I.~N., Sokolovsky, K.~V., \& Gavras, P.\ 2018, \mnras, 475, 2326 

\bibitem[Pearson 1901]{pearson1901} Pearson, K.\ 1901, Philosophical Magazine 2 (11): 559–572

\bibitem[Pearson et al. 2018]{pearson2018} Pearson, K.~A., Palafox, L., \& Griffith, C.~A.\ 2018, \mnras, 474, 478 

\bibitem[Pilbratt et al. 2010]{pilbratt2010} Pilbratt, G.~L., Riedinger, J.~R., Passvogel, T., et al.\ 2010, \aap, 518, L1 

\bibitem[Poglitsch et al. 2010]{poglitsch2010} Poglitsch, A., Waelkens, C., Geis, N., et al.\ 2010, \aap, 518, L2 

%\bibitem[Provost \& Fawcett 2013]{provost2013} Provost, F., \& Fawcett, T. 2013, \textit{Data Science for Business}, First Edition, 
%O'Reilly Media, Inc.

\bibitem[Quinlan 1986]{quinlan1986} Quinlan, J.~R.\ 1986, Induction of Decision Trees, Machine Learning 1: 81-106, 
Kluwer Academic Publishers

\bibitem[Rathborne et al. 2010]{rathborne2010} Rathborne, J.~M., Jackson, J.~M., Chambers, E.~T., 
et al.\ 2010, \apj, 715, 310 

\bibitem[Rieke et al. 2004]{rieke2004} Rieke, G.~H., Young, E.~T., Engelbracht, C.~W., et al.\ 2004, \apjs, 154, 25 

\bibitem[Rosenblatt 1958]{rosenblatt1958} Rosenblatt, F.\ 1958, Psychological Review, 65 (6): 386-408

\bibitem[Saar-Tsechansky \& Provost 2007]{saar2007} Saar-Tsechansky, M. \& Provost, F. 2007, Journal of Machine Learning Research, 8, 1625-1657

\bibitem[Sauvage et al. 2014]{sauvage2014} Sauvage, M., Okumura, K., Klaas, U., et al.\ 2014, Experimental Astronomy, 37, 397 

\bibitem[Scholkopf \& Smola 2001]{scholkopf2001} Scholkopf, B., \& Smola, A.~J.\ 2001, 
\textit{Learning with kernels: support vector machines,
regularization, optimization, and beyond}, MIT press

\bibitem[Shetty et al. 2009]{shetty2009} Shetty, R., Kauffmann, J., Schnee, S., Goodman, A.~A., \& Ercolano, B.\ 2009, \apj, 696, 2234 

\bibitem[Siringo et al. 2009]{siringo2009} Siringo, G., Kreysa, E., Kov{\'a}cs, A., et al.\ 2009, \aap, 497, 945 

\bibitem[Siringo et al. 2010]{siringo2010} Siringo, G., Kreysa, E., De Breuck, C., et al.\ 2010, The Messenger, Vol.~139, p.~20 

\bibitem[Skrutskie et al. 2006]{skrutskie2006} Skrutskie, M.~F., Cutri, R.~M., Stiening, R., et al.\ 2006, \aj, 131, 1163 

\bibitem[Spezzi et al. 2015]{spezzi2015} Spezzi, L., Petr-Gotzens, M.~G., Alcal{\'a}, J.~M., et al.\ 2015, \aap, 581, A140 

\bibitem[Sreejith et al. 2018]{sreejith2018} Sreejith, S., Pereverzyev, S., Jr., Kelvin, L.~S., et al.\ 2018, \mnras, 474, 5232 

\bibitem[Stutz et al. 2013]{stutz2013} Stutz, A.~M., Tobin, J.~J., Stanke, T., et al.\ 2013, \apj, 767, 36 

\bibitem[Tangaro et al. 2015]{tangaro2015} Tangaro, S., Amoroso, N., Brescia, M., et al.\ 2015,  
Computational and Mathematical Methods in Medicine, Vol.~2015, Article ID 814104

%\bibitem[Vapnik 2000]{vapnik2000} Vapnik, V. 2000, \textit{The Nature of Statistical Learning Theory}, Second Edition, Springer-Verlag New York

\bibitem[Vapnik \& Lerner 1963]{vapnik1963} Vapnik, V., \& Lerner, A. 1963, Automation and Remote Control, 24, 774–780

\bibitem[van Buuren \& Groothuis-Oudshoorn 2011]{vanbuuren2011} van Buuren, S. \& Groothuis-Oudshoorn, K.\ 2011, Journal of Statistical Software, 45(3), 1-67

\bibitem[White et al. 2007]{white2007} White, R.~J., Greene, T.~P., Doppmann, G.~W., et al.\ 2007, \textit{Protostars and Planets V},
B. Reipurth, D. Jewitt, and K. Keil (eds.), University of Arizona Press, Tucson, 951 pp., p.~117 

\bibitem[Witten \& Fran 2005]{witten2005} Witten, I.~H., \& Frank, E. 2005, \textit{Data Mining - Practical Machine Learning Tools and 
Techniques}, Second Edition, Morgan Kaufmann Publishers, Elsevier Inc.

\bibitem[Yan et al. 2018]{yan2018} Yan, Q.-Z., Xu, Y., Walsh, A.~J., et al.\ 2018, \mnras, 476, 3981 

\bibitem[Ybarra \& Lada 2009]{ybarra2009} Ybarra, J.~E., \& Lada, E.~A.\ 2009, \apjl, 695, L120 

\bibitem[Yee et al. 2017]{yee2017} Yee, J.~C., Fazio, G.~G., Benjamin, R., et al.\ 2017, {\tt arXiv:1710.04194} 

\bibitem[Zhang 2000]{zhang2000} Zhang, G.\ 2000, IEEE Trans Syst Man Cy C 30(4):451-462

\bibitem[Zhanq 2004]{zhang2004} Zhang, H.\ 2004, AA, 1(2), 3


\end{thebibliography}

%% Non-BibTeX  (Name-Year style)
%

\end{document}